%% file: FENNI_Part_II_Arxiv.tex
\documentclass{article}

\usepackage{Arxiv}

\usepackage{MyNotations}

\usepackage{amssymb}

\usepackage[backend=biber,bibstyle=custom-apa,labeldate=year,citestyle=apa,isbn=false,url=false,eprint=false,backref=true,sortcites=false,sorting=nyt]{biblatex} 
\newcommand{\mkbibbracketscol}[1]{\mkbibbrackets{\textcolor{accentcolor!80}{#1}}}
\DeclareCiteCommand{\parencite}[\mkbibbracketscol]
{\usebibmacro{cite:init}%
	\usebibmacro{prenote}%
	\toggletrue{apa:inpcite}}
{\usebibmacro{citeindex}%
	\printtext[bibhyperref]{\usebibmacro{cite}%
		\usebibmacro{cite:post}%
		\togglefalse{apa:inpcite}}}
{\multicitedelim}
{\usebibmacro{postnote}}

\usepackage{microtype}
\usepackage{doi}

\title{Finite Element Neural Network Interpolation. \\ Part II: Hybridisation with the Proper Generalised Decomposition for non-linear surrogate modelling}

\author{
	Alexandre Daby-Seesaram \thanks{Corresponding author} \\
	LMS, École Polytechnique,\\
	 IPP/CNRS, Palaiseau, France.\\
	   \texttt{} \\
	INRIA, Palaiseau, France.\\
	\href{mailto:alexandre.daby-seesaram@polytechnique.edu}{\texttt{alexandre.daby-seesaram@polytechnique.edu}} \\
	\And
Kateřina Škardová \\
	LMS, École Polytechnique,\\
IPP/CNRS, Palaiseau, France.\\
  \texttt{} \\
INRIA, Palaiseau, France.\\
\href{mailto:katerina.skardova@polytechnique.edu}{\texttt{katerina.skardova@polytechnique.edu}} \\
	\And
Martin Genet \\
	LMS, École Polytechnique,\\
IPP/CNRS, Palaiseau, France.\\
  \texttt{} \\
INRIA, Palaiseau, France.\\
\href{mailto:martin.genet@polytechnique.edu}{\texttt{martin.genet@polytechnique.edu}} \\
}

\usepackage{tocloft}  

\begin{document}

		\maketitle

		\begin{abstract}
			This work introduces a hybrid approach that combines the Proper Generalised Decomposition (PGD) with deep learning techniques to provide real-time solutions for parametrised mechanics problems. By relying on a tensor decomposition, the proposed method addresses the curse of dimensionality in parametric computations, enabling efficient handling of high-dimensional problems across multiple physics and configurations. Each mode in the tensor decomposition is generated by a sparse neural network within the Finite Element Neural Network Interpolation (FENNI) framework presented in Part I, where network parameters are constrained to replicate the classical shape functions used in the Finite Element Method. This constraint enhances the interpretability of the model, facilitating transfer learning, which improves significantly the robustness and cost of the training process. The FENNI framework also enables finding the optimal spatial and parametric discretisation dynamically during training, which accounts to optimising the model's architecture on the fly. This hybrid framework offers a flexible and interpretable solution for real-time surrogate modelling. We highlight the efficiency of the FENNI-PGD approach through 1D and 2D benchmark problems, validating its performance against analytical and numerical reference solutions. The framework is illustrated through linear and non-linear elasticity problems, showing the flexibility of the method in terms of changes in physics.
		\end{abstract}


	\newpage
	
	\section{Introduction}
	
	In many engineering fields, having a digital twin of a structure or object is of great interest in understanding and predicting its behaviour throughout its lifetime. In a medical context, for instance, having a digital twin of a patient's organ can help with personalised medical diagnosis or prognosis \parencite{patteEstimationRegionalPulmonary2022, lavilleComparisonOptimizationParametrizations2023,laubenbacherDigitalTwinsMedicine2024, gonsardDigitalTwinsChronic2024}. To do so, the digital twin must provide a near real-time solution to the mechanics problem so clinicians can draw conclusions during the consultation. Additionally, rapid access to solutions is crucial for conducting the numerous simulations necessary for the model identification phase required to develop personalised digital twins. 
	For the digital twin to be clinically relevant, the organ's simulation requires combining different computations (possibly involving different physics) at multiple scales \parencite{burrowesVirtualLungMultiscale2008,leeMultiphysicsComputationalModeling2016,chabiniokMultiphysicsMultiscaleModelling2016}. The wide variety of problems that need to be accounted for requires the chosen framework to be robust and versatile, and able to handle non-linear problems.
	
	\bigbreak
	
	Regarding the real-time requirement of the digital twin, several reduced-order modelling (ROM) techniques have been developed to reduce the computational burden associated with parametric problems. Most of these methods require solving high-fidelity models for selected parameter sets, referred to as \emph{snapshots}. These \emph{snapshots} are then compressed into a reduced-order basis, onto which the original problem is projected. This projection leads to a low-dimensional system that can be solved more efficiently for new parameters. One main example of this approach is the Proper Orthogonal Decomposition (POD) \parencite{sirovichTurbulenceDynamicsCoherent1987,chatterjeeIntroductionProperOrthogonal2000}, which significantly reduces the computational burden of parametric simulations. However, the quality of the low-rank approximation strongly depends on the choice of snapshots. To remove this arbitrariness, the reduced-basis method \parencite{madayPrioriConvergenceTheory2002,madayReducedBasisElementMethod2002,quarteroniCertifiedReducedBasis2011,quarteroniReducedBasisMethods2016} offers an automatic strategy for selecting snapshots, minimising the need for numerous full-order computations. Unlike these \emph{a posteriori} techniques, \emph{a priori} methods such as the Proper Generalised Decomposition (PGD) \parencite{chinestaShortReviewModel2011} construct the reduced-order basis dynamically without requiring prior high-fidelity solutions. Initially introduced for space-time problems under the name \emph{radial approximation} \parencite{ladevezeNouvelleStrategieCalcul1999}, the PGD has since been extended to more general parametric scenarios \parencite{luMultiparametricSpacetimeComputational2018, niroomandiRealtimeSimulationBiological2013,neronTimespacePGDRapid2015,daby-seesaramHybridFrequencytemporalReducedorder2023}. In some cases, the PGD greedy construction is referred to as the ``offline phase'', where the model is built. Once constructed, the ``online phase'' only involves evaluating the surrogate model for new parameter sets, eliminating the need for further equation solving \parencite{niroomandiRealtimeSimulationBiological2013}. Unlike \emph{a posteriori} approaches, the \emph{a priori} framework enables surrogate modelling, thus eliminating the need for online computations; the online phase is reduced to evaluating the model for specific parameters. Moreover, the PGD is not limited to offering a surrogate for a specific quantity of interest but also enables straightforward evaluation of the full mechanical response across the entire structure. As a result, these tools enable accounting for high-dimensional problems while avoiding the curse of dimensionality. The PGD, for instance, leads to a problem's complexity that scales linearly with the number of parameters \parencite{luAdaptiveSparseGrid2018}. 
	
	As for the robustness and versatility requirement of the chosen reduced-order modelling framework for tackling the wide variety of problems involved in finding a digital twin at the organ's scale, the tools used in deep learning frameworks offer promising prospects. Indeed, finding a PGD solution that actually minimises the error is generally complex and problem-dependent, rendering the task of having a unified PGD framework difficult \parencite{nouyPrioriModelReduction2010}. However, the increasing popularity of Physics-informed Neural Networks (PINNs), for instance, highlights that using deep-learning libraries allows constructing surrogate models for various problems with relative ease. Since the original proposition of the PINNs \parencite{raissiPhysicsinformedNeuralNetworks2019a}, many publications have explored using PINNs for surrogate modelling across various applications. In \parencite{haghighatPhysicsinformedDeepLearning2021}, the authors use PINNs for surrogate modelling in solid mechanics, demonstrating how multiple separate networks connected using the constitutive relations can be used to solve parametric PDEs by training the PINN-based model on a grid of parameter values. Similarly, surrogate PINNs have been used to enable fast inverse problem-solving in a biomedical context \parencite{caforioPhysicsinformedNeuralNetwork2024}. Surrogate models based on Graph Neural Networks (GNN) have also been proposed for medical applications, enabling fast evaluation of patient-specific parametrised solutions \parencite{daltonPhysicsinformedGraphNeural2023}. Such methods appear versatile and relatively accurate at the cost of many trainable parameters. In an effort to reduce the number of trainable parameters, several approaches propose using deep learning paired with a reduced-order basis (ROB). PINNs have, for instance, been used to replace the projection phase of a POD approach \parencite{hijaziPODGalerkinReducedOrder2023}, where neural networks are used to learn the coefficient associated with space modes obtained through the compression of snapshots. Another approach consists of using a (Graph) neural network to learn the reduced-order basis during supervised training. The modes can then be evaluated for several topologies and serve as an initial ROB when using the PGD \parencite{matrayHybridNumericalMethodology2024a}.
	
	However, for these PINN-based methods, changing the discretisation of the mesh used for the interpolation is not straightforward and often leads to the need to re-train the model from scratch. 
	The idea to use FEM interpolation to compute the loss has been recently introduced \parencite{duNeuralIntegratedMeshfreeNIM2024,motiwaleNeuralNetworkFinite2024,badiaFiniteElementInterpolated2024}. While it helps with getting exact values of the space integrals as well as strongly prescribing Dirichlet boundary conditions, the interpretability of such networks remains difficult. On the other hand, in the Embedded Finite Element Neural Networks (EFENN) introduced with the Hierarchical Deep Learning Neural Networks (HiDeNN) \parencite{zhangHierarchicalDeeplearningNeural2021}, shape functions are generated directly within the neural network architecture, making it sparse and fully determined by the problem's discretisation. This approach leads to interpretable model parameters and enables simultaneously optimising nodal coordinates and values, offering the potential for mesh adaptivity and transfer learning. This approach has been used to provide an automatic Finite Element Neural Network Interpolation (FENNI) method in which the interpolation is optimised during the training stage to provide a solution minimising, among other things,  the potential energy of the system \parencite{skardovaFiniteElementNeural2024}. 
	
	\bigbreak
	
	In this paper, we propose to hybridise deep learning tools and standard reduced-order modelling techniques. This method aims to bring the ease of use of the PINNs to PGD by building on the fly a tensor decomposition whose modes are generated by a sparse neural network within the Finite Element Neural Network Interpolation (FENNI) framework. We propose a new way to find the PGD tensor decomposition based on back-propagation and optimisers used in machine learning applications. This allows us to build a unified way of solving reduced-order problems, thus circumventing the difficulty of a problem-dependent implementation classically encountered with the standard PGD algorithms. This method also offers greater versatility in calculating modes than conventional sequential greedy methods. This versatility allows several linear or non-linear physics to be easily tackled with minimal changes. The proposed framework also allows the spatial or parametric mesh to be adapted on the fly during the computation, leading to a fine solution while controlling the required number of degrees of freedom.
	Finally, unlike most approaches based on neural networks, the interpretability offered by the FENNI framework enables transfer learning between different models with different architectures, leading to high efficiency in the model's design and limiting the wasteful use of resources. The model's architecture is indeed optimised on the fly during the training stage. 
	
	\cref{sec:pb_set} presents the mechanical problem used to test the proposed framework and briefly recalls the standard interpolation methods for the numerical solution. This section also summarises the recently introduced FENNI framework, which has been further detailed in Part I \parencite{skardovaFiniteElementNeural2024}. \cref{sec:method} describes the surrogate models and the hybridisation between the FENNI framework and the PGD, detailing how the tensor decomposition is built on the fly during the model's training. Numerical examples are shown in \cref{sec:Num}.
	
	\section{Problem setting}
	\label{sec:pb_set}
	\subsection{Reference problem}
	\label{sec:Ref_pb}
\begin{figure}[H]
		\centering
		\includegraphics[width=0.25\linewidth]{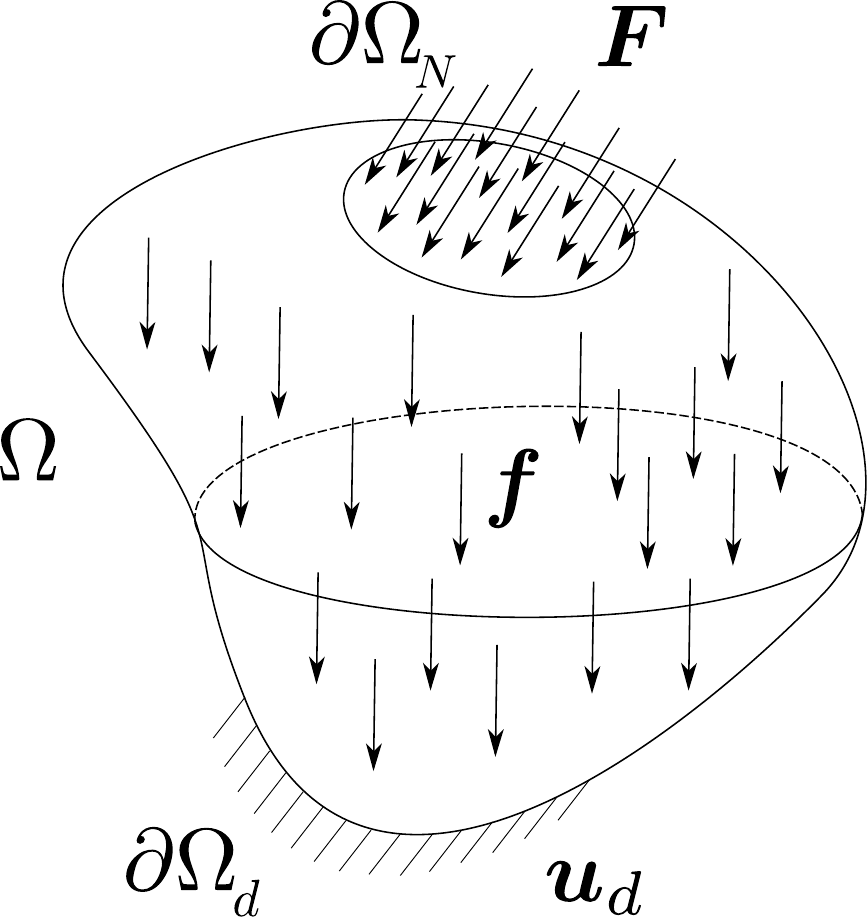}
		\caption{Reference problem}
		\label{fig:ref}
	\end{figure}

	The surrogate modelling capabilities of the method are illustrated on static elasticity problems. Let us consider a body described by a bounded domain $\Omega$ submitted to body forces $\vect{f}$ and surface forces $\vect{F}$  on $\Omega$ and $\dSn$, respectively, as represented in \cref{fig:ref}. Displacements $\vect{u}_d$ are prescribed on $\dSd$.  To account for various patients and situations, the material and loading variability is represented by a set of $\beta$ extra parameters $\para$. 
	
	When considering only conservative forces, solving the mechanical problem means finding a displacement field $\vect{u}$ that minimises the potential energy
	\begin{equation}
		E_p\left(\para\right) := {E_p}_{int}\left(\para\right) + {E_p}_{ext} \left(\para\right)
		\label{eq:MechPb}
	\end{equation}
	that also satisfies the Dirichlet boundary conditions
	
	\begin{equation}
			\vect{u} = \vect{u}_d \qquad \qquad \text{on } \partial \Omega_{d}. 
	\end{equation}\hfill

	${E_p}_{int}\left(\para\right)$ is the strain energy, whose expression depends on the chosen behaviour and 
	\begin{equation}
		{E_p}_{ext}  = \intSn - \vect{F}\cdot \vect{u} \dS + \intV - \vect{f}\cdot\vect{u}\dV
	\end{equation}
	is the potential energy of the external forces applied onto the structure.
	%

	For each snapshot, \emph{i.e.} for each set of parameters, solving the problem involves finding the displacement solution within the kinematically admissible (resp. admissible to zero) displacement space
	
	\begin{equation}
		\mathcal{U} \left(\text{resp.} \mathcal{U}^0 \right) = \left\{\vect{u} \; | \; \vect{u}\left(\vect{x}\right) \in \mathcal{H}^1\left(\Omega, \mathbb{R}^{\text{d}}\right) \text{, }
		\vect{u} = \vect{u}_d \left(\text{resp.} =\vect{0} \right) \text{ on }\partial \Omega_d  \right\}  \text{,}
	\end{equation}
	where $\text{d} \in \left\{1,2,3\right\}$ is the dimension of the problem. In turn, the parametrised displacement field  $\vect{u}\left(\vect{x},\para \right)$ lies in the admissible (resp admissible to zero) space  			
	\begin{equation}
		\mathcal{U}^{\beta} \left(\text{resp.} \mathcal{U}^{0,~\beta} \right) = \left\{\vect{u} \; | \; \vect{u}\left(\vect{x},\para\right) \in \mathcal{H}^1\left(\Omega, \mathbb{R}^{\text{d}}\right) \otimes \mathcal{L}^2\left(\mathcal{B}, \mathbb{R}^{\beta}\right) \text{, }
		\vect{u} = \vect{u}_d \left(\text{resp.} =\vect{0} \right) \text{ on }\partial \Omega_d  \right\}  \text{,}
	\end{equation}
	where $\mathcal{B}$ is the parameters hypercube. 
	
	\subsection{Solution interpolation: FEM, FENNI framework and Tensor Decomposition }
	\label{sec:Interpolation}
	In order to solve the mechanical problem numerically, the admissible space of infinite dimensions is discretised into a space of possibly large but finite dimensions. 
	
	\subsubsection{Recall of Galerkin method}
	Solving a mechanics problem using the Finite Element Method (FEM) implies looking for the solution as a linear combination of $N$ shape functions, $N$ being the number of degrees of freedom (Dofs). For $0$-form shape functions, for instance, $N = n_p \times d$, with $n_p$ being the number of nodes and $d$ the dimension of the problem.
	Instead of looking for the exact solution within the admissible displacement space $\mathcal{U}$
	an approximation of the solution is sought after in the finite discrete admissible (resp. admissible to zero) space (of dimension $N$)
	\begin{equation}
		\mathcal{U}_h \left(\text{resp.} \mathcal{U}_h^0 \right) = \left\{\vect{u}_h  \; | \; \vect{u}_h \in \text{Span}\left( \left\{ N_i\left(\vect{x} \right)\right\}_{i \in \llbracket 1,n_p\rrbracket} \right)^d \text{, }
		\vect{u}_h = \vect{u}_d \left(\text{resp.} =\vect{0} \right) \text{ on }\partial \Omega_d  \right\}  \text{.}
	\end{equation}
	
	In the case of a parametric solution field, instead of searching the displacement field $\vect{u}\left(\vect{x},\para \right)$, for instance, in the admissible solution space $\mathcal{U}^{\beta}$,
	an approximation of the solution is sought after in a separate variable form in the semi-discrete space
	\begin{equation}
		\mathcal{U}_h^{\beta} \left(\text{resp.} \mathcal{U}_h^{0,\beta} \right) = \left\{\vect{u}_h  \; | \; \vect{u}_h\left(\vect{x},\para\right) \in \text{Span}\left( \left\{ N_i\left(\vect{x} \right)\right\}_{i \in \llbracket 1,n_p\rrbracket} \right)^d\otimes \mathcal{L}^2\left(\mathcal{B}, \mathbb{R}^{\beta}\right) \text{, }
		\vect{u}_h = \vect{u}_d \left(\text{resp.} =\vect{0} \right) \text{ on }\partial \Omega_d  \right\}  \text{,}
	\end{equation}
	leading to 
	\begin{equation}
		\vect{u} \simeq\vect{u}_h\left(\textcolor{BleuLMS!70}{\vect{x}}, \textcolor{LGreenLMS}{\para}\right) = \sum\limits_{i=1}^N \textcolor{BleuLMS!70}{\vect{N}_i(\vect{x})} ~\textcolor{LGreenLMS}{\lambda_i\left( \para \right)}.
		\label{eq:FEM}
	\end{equation}
	In turn, the parametric mapping $\mathcal{L}^2\left(\mathcal{B}, \mathbb{R}^{\beta}\right) $ can be discretised. 
	The FEM method then consists in looking for the solution in a separated-variable form in order to change the problem from finding the solution in an infinite space to a finite space, relying on the Galerkin method. To reduce the computational complexity of the solved FEM problems, a reduced-order tensor decomposition can be applied, further lowering the dimensionality of the solution space.
	
	\subsubsection{Recall on the Finite Element Interpolation Neural Network (FENNI) framework}

	Recent research has aimed to integrate Finite Element Methods (FEM) with neural networks, yielding solutions that, unlike traditional PINNs, are based on pre-defined shape functions. Although these methods have limited approximation flexibility compared to fully connected neural networks, they offer increased control and interpretability, especially in handling boundary conditions. 
	
	The Neural-integrated meshfree (NIM) method proposed by \cite{duNeuralIntegratedMeshfreeNIM2024} employs mesh-free basis functions associated with nodal coefficients predicted by neural networks, while Dirichlet boundaries are managed through additional loss terms. 
	
	In contrast, in VPINNS \parencite{badiaFiniteElementInterpolated2024} the output of a fully-connected neural network is projected onto FEM shape functions. The training is performed using a loss function based on a weak form of the PDE residuals. Prescribing Dirichlet boundary conditions is more straightforward than with PINNs, as they can be strongly prescribed after the projection to the shape function space.
	
	The HiDeNN method \parencite{zhangHierarchicalDeeplearningNeural2021} directly generates shape functions within the neural network, enabling all interpolation parameters to be interpretable and optimised together. This approach is the starting point of the FENNI framework presented in Part I \parencite{skardovaFiniteElementNeural2024}, where \emph{hr-adaptivity} is inherently incorporated into the interpolation of the solution. The interpretability of the model, arising from the nodal values and coordinates being its parameters, allows Dirichlet boundary conditions to be easily prescribed by fixing specific nodal values. Moreover, the simultaneous training of both nodal values and coordinates enables efficient mesh adaptivity. The architecture of the FENNI model is determined by the discretisation of the mesh and is dynamically optimised by relying on the high transfer-learning capabilities offered by the interpretability of the framework.
	
	\subsubsection{Recall on the Tensor Decomposition}
	As can be seen in \cref{eq:FEM}, the FEM interpolation already relies on the separation of variables. In the context of reduced-order modelling, shape functions can be viewed as highly localised space modes with compact support that are easily defined. The objective of reduced-order modelling is to find fewer global modes that would replace the shape functions in the solution interpolation. 
	Using $m$ global modes in the linear combination instead of the original $N$ can decrease the number of Dofs, \emph{i.e.,} the solution can be well approximated using $m \ll N$ modes. Doing so means looking for a solution in a much smaller, reduced space spanned by the global modes. In this framework, the reduced-order expression of  a displacement field $\vect{u}$ parameterised by the position $\vect{x}$ and a set of $\beta$ parameters $\para$ using  $m$ modes reads
	\begin{equation}
		\vect{u}\left(\textcolor{BleuLMS!70}{\vect{x}}, \textcolor{LGreenLMS}{\para}\right) = \sum\limits_{i=1}^m \textcolor{BleuLMS!70}{\overline{\vect{u}}_i(\vect{x})} ~\textcolor{LGreenLMS}{\prod_{j=1}^{\beta}\lambda_i^j(\mu_j)} \in \mathcal{U}_{\text{ROM}},
		\label{eq:ND_PGD}
	\end{equation}
	with 
	\begin{equation}
		\mathcal{U}_{\text{ROM}} \left(\text{resp.} \mathcal{U}_{\text{ROM}}^{0} \right) = \left\{\vect{u} \; | \; \vect{u}\left(\vect{x},\para\right) \in \mathcal{U}_h \bigotimes_{i=1}^{\beta} \mathcal{L}^2\left(\mathcal{B}_i, \mathbb{R}\right) \text{, }
		\vect{u} = \vect{u}_d \left(\text{resp.} =\vect{0} \right) \text{ on }\partial \Omega_d  \right\}  \text{,}
	\end{equation}
	Note that the tensor decomposition also occurs in the parameter space, which, contrary to the FEM interpolation, is also represented as the tensor product of several reduced dimension spaces.
	
	Using \emph{a priori} reduced-order modelling methods such as the PGD removes the need for prior full-order computations in order to build the ROB from snapshots. The modes are instead built on the fly during the computations. 
	In order to account for non-homogeneous Dirichlet boundary conditions, the solution can be written as a sum of admissible and homogeneous (admissible to zero) fields. To do so, the first space mode is set with the actual boundary conditions, while the following modes are set up with homogeneous boundary conditions. To avoid interfering with the space boundary conditions, the parametric modes associated with the first modes are uniformly equal to one. The displacement thus reads
	\begin{equation}
		\vect{u}\left(\textcolor{BleuLMS!70}{\vect{x}}, \textcolor{LGreenLMS}{\para}\right) = \underbrace{\textcolor{BleuLMS!70}{\overline{\vect{u}}_1(\vect{x})} ~\textcolor{LGreenLMS}{\prod_{j=1}^{\beta}\lambda_1^j(\mu_j)} }_{\text{Admissible field}}+ \sum\limits_{i=2}^m \underbrace{\textcolor{BleuLMS!70}{\overline{\vect{u}}_i(\vect{x})} ~\textcolor{LGreenLMS}{\prod_{j=1}^{\beta}\lambda_i^j(\mu_j)}}_{\text{Homogeneous corrections}}.
	\end{equation}
	\begin{equation}
		\begin{cases}
			\lambda_1^j(\mu_j) = 1, \forall j \in \llbracket 1,\beta \rrbracket \\[7pt]
			\overline{\vect{u}}_1(\vect{x}) = \vect{u}_d \text{ on } \partial \Omega_d \\[7pt]
			\overline{\vect{u}}_i(\vect{x}) = \vect{0} 
			\text{ on } \partial \Omega_d, \forall i  \in \llbracket 2,m \rrbracket
		\end{cases}
	\end{equation}
	The first mode is thus kinematically admissible, while the following modes are admissible to zero.
	To find the appropriate continuous space and parameter modes, the following minimisation problem must be solved
	\begin{equation}
		\left(\left\{\overline{\vect{u}}_i \right\}_{i\in \llbracket 1,m\rrbracket},\left\{\lambda_i^j \right\}_{i\in \llbracket 2,m\rrbracket}^{j\in \llbracket 1,\beta\rrbracket} \right) = \argmin_{
			\left\{
			\begin{array}{l}
				\left(\overline{\vect{u}}_1, \left\{\overline{\vect{u}}_i \right\}_{i\in \llbracket 2,m\rrbracket} \right) \in \mathcal{U}\times \mathcal{U}_0 \\[4mm] 
				\left\{\lambda_i^j \right\}_{i\in \llbracket 2,m\rrbracket}^{j\in \llbracket 1,\beta\rrbracket} \in \left( \bigtimes_{j=1}^{~\beta} \mathcal{L}_2\left(\mathcal{B}_j\right) \right)^{m-1}
			\end{array}
			\right\}
		} \int_{\mathcal{B}} E_p\left(\vect{u}\right) \, \mathrm{d} \beta,
		\label{eq:min_problem}
	\end{equation}
	where the $\mathcal{B}_i$ are defined such that $\mathcal{B} = \bigcup_{i=1}^{\beta} \mathcal{B}_i$.
	
	Standard algorithms proposed to solve this minimisation problem are problem-dependent and require solving an adjoint problem \parencite{nouyPrioriModelReduction2010}. This makes it challenging to write a solver that can handle a wide number of problems. In practice, instead of solving the minimisation problem, several studies rely on a  (Petrov-)Galerkin PGD, which offers an easier implementation at the cost of a non-monotonous residual convergence. The use of optimisers described in \cref{sec:training} enables straightforwardly minimising any loss function for a wide variety of physical losses.  
	Relying on such optimisers also allows for an easy implementation of algorithms that would perform the minimisation on several modes instead of simply optimising the new modes during the greedy process of building the reduced-order basis.

	\section{Methods}
	\label{sec:method}
	\subsection{Proposal: leveraging the FENNI framework in the tensor decomposition}
	
	We aim to build a unified framework that would make the concept of PGD easily applicable to a wide range of problems. We propose to combine the concept of tensor decomposition with the Finite Element Neural Network Interpolation framework, utilising the using the \emph{Multiplication block} $\mathcal{M}$ detailed in \parencite{zhangHierarchicalDeeplearningNeural2021}. 
	Specifically, we will be using a framework with an alternative reference-element-based architecture and extended mesh adaptivity capabilities: Finite Element Neural Network Interpolation (FENNI). The method's efficiency in solving PDEs on adaptive meshes was demonstrated in \parencite{skardovaFiniteElementNeural2024}. 	The PGD built using the FENNI framework (FENNI-PGD) will leverage all the advantages of this framework. Doing so answers the versatility requirement of the chosen framework. Indeed, the only component that changes based on the governing equations is the loss function. The framework is, therefore, universal and can be used for a variety of problems.
	
	Such an implementation falls in the scope of \emph{a priori} reduced-order modelling \parencite{ryckelynckPrioriModelReduction2006,chinestaShortReviewModel2011} as no prior knowledge is required and the method does not require any full-order computations to construct a reduced order basis.
	The FENNI-PGD proposed in this paper is based on the HiDeNN tensor decomposition proposed in \parencite{zhangHiDeNNTDReducedorderHierarchical2022} but built \emph{greedily} and used for surrogate modelling. In the proposed FENNI-PGD, the number of modes $m$ is not fixed in advance and evolves during the training stage until convergence is reached. The details of the mode addition, as well as the convergence criterion, are further detailed in the next section.  
	
	The FENNI-PGD is built as a combination of several FENNI networks \parencite{skardovaFiniteElementNeural2024}, each representing a mode in the tensor decomposition. The output is then given as the tensor product between all the modes. In a context where $\beta = 1$, \emph{i.e.} there is a single extra coordinate in the tensor decomposition, there are $m$ FENNI networks for the parametric modes and $m$ FENNI networks for the interpolation of the space modes.
	\begin{equation}
		\vect{u}(\textcolor{BleuLMS!70}{\vect{x}},\textcolor{LGreenLMS}{\mu}) = \sum\limits_{i=1}^m \textcolor{BleuLMS!70}{\overline{\vect{u}}_i(\vect{x})} ~\textcolor{LGreenLMS}{\lambda_i(\mu)}
	\end{equation}

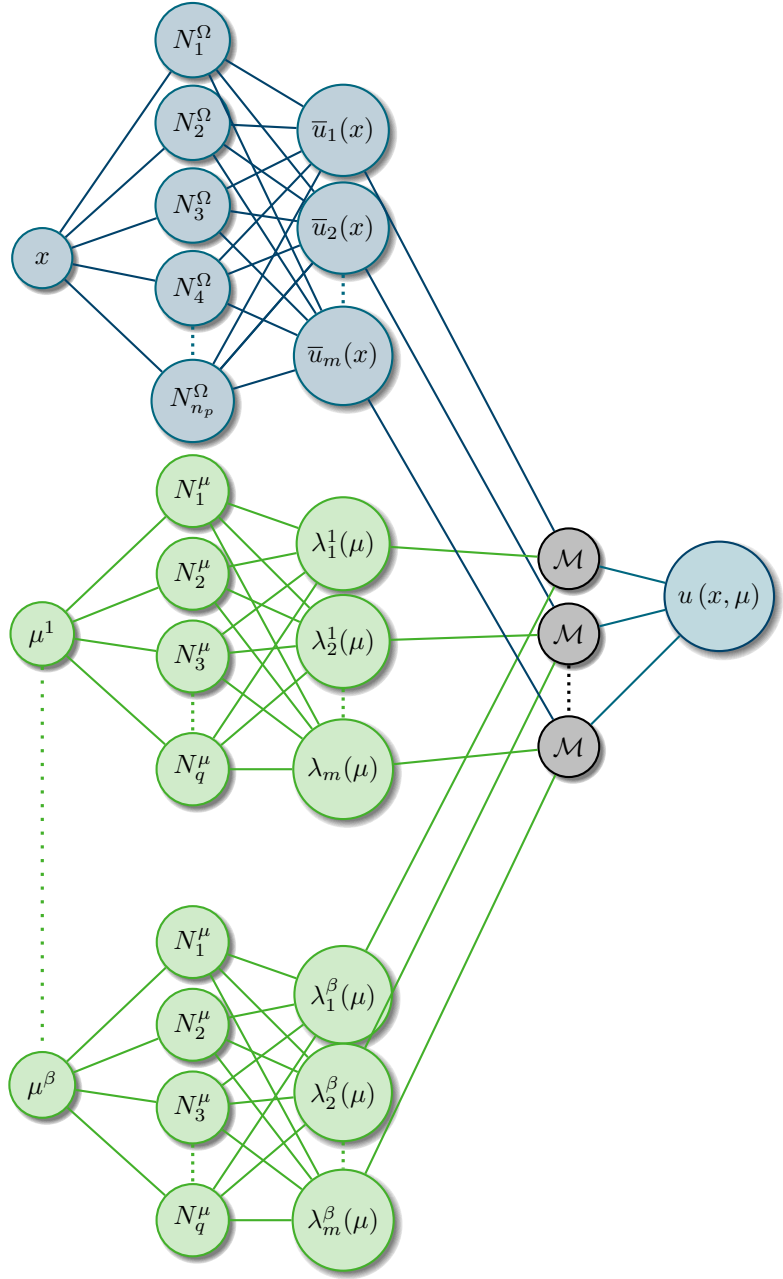
\begin{figure}[H]
		\centering
		\input{FENNI_PGD.tex}
		\caption{FENNI-PGD interpolation for multiple parameters $\left\{\mu_i\right\}_{i \in \llbracket 1,\beta\rrbracket}$}
		\label{fig:HiDeNN_TD_multipleMu}
	\end{figure}

	The FENNI-PGD representation can be extended to any number of parameters, \emph{i.e.} considering $\beta > 1$. In such a case, there are $m$ FENNI networks for the interpolation of the space modes and $\beta \times m$ FENNI networks for the parametric modes as shown in \cref{fig:HiDeNN_TD_multipleMu}. This architecture corresponds to the tensor decomposition given in \cref{eq:ND_PGD}.

	\subsection{The FENNI-PGD: computing the tensor decomposition \emph{aka} the training phase}
	\label{sec:training}

	Solving the mechanical problem described in \cref{sec:Ref_pb} amounts to minimising the associated potential energy defined in  \cref{eq:MechPb}.
	In a parametric context, the loss function to to be minimised is defined as the potential energy averaged over the parametric space.
	\begin{equation}
		\mathcal{L} := \overline{E_p}\left(\vect{u}\left(\vect{x},\para\right),\para\right) =  \int_{\mathcal{B}}E_p\left(\vect{u},\para\right)\mathrm{d}\beta.
		\label{eq:loss}
	\end{equation}
	The implementation of the different losses used in this paper (linear elasticity and finite strain elasticity) is detailed in \cref{ap:loss}. As previously mentioned, the tensor decomposition is not fixed at the beginning of the training stage but evolves during the computations. The modes are added on the fly until convergence is reached. 
	Contrary to most PGD algorithms, the $m-1$ previously added modes continue to be trained once the $m$-th mode is added. This provided more flexibility in the tensor decomposition to minimise the loss function, thus improving the overall optimality of the TD.

	\subsubsection{Greedy algorithm}
	
	The tensor decomposition modes are added in a greedy manner during the training process. The convergence criterion relies on the stagnation of the loss given by the decay of the loss gradient with regard to the epoch number. In other word, stagnation is reached when 
	\begin{equation}
		\hat{\mathcal{L}} := \frac{\nabla_{\text{epoch}} \mathcal{L}}{\mathcal{L}} =2 \frac{\mathcal{L}_{n-1} - \mathcal{L}_n}{\mathcal{L}_{n-1} + \mathcal{L}_n} \le \eta_c,
	\end{equation}
	with $\eta_c$ a stagnation threshold and $\mathcal{L}_n$ the value of the loss at the $n\text{-th}$ iteration. When stagnation is reached, a new mode is added. If the loss keeps stagnating then the training is considered over and the tensor decomposition is converged, otherwise the training continues. The resulting greedy algorithm is summarised in Algorithm \ref{alg:GreedyPGD}.
	The hyperparameters \code{\arbitrary{max\_stgn}} and \code{\arbitrary{new\_mode\_threshold}} were not shown to significantly affect the algorithm's performance, even when chosen within a broad range (5 to 100). This suggests that the algorithm is not overly sensitive to the specific values of these two hyperparameters, allowing for flexibility in their selection without compromising the overall behaviour of the model.
	The convergence precision is driven by the convergence criterion \code{\arbitrary{$\eta_c$}}.

\input{Algo_GreedyPGD}

	\subsubsection{Two-Stage optimisation}

	The minimisation problem described in \cref{eq:min_problem} is solved using variants of the gradient descent algorithm. Gradient descent is a well-established optimisation method used to minimise objective functions by iteratively adjusting their parameters. Recently, variants of this method adapted for training neural networks have been developed.
	More specifically, a two-stage training process is employed, combining the Adam and L-BFGS optimisers. In the first stage, Adam, a variant of stochastic gradient descent, is used to explore the parameter landscape. By leveraging momentum, Adam \parencite{kingmaAdamMethodStochastic2014} helps to avoid local minima and identify a suitable initialisation region for the second stage. In the second stage, L-BFGS \parencite{liuLimitedMemoryBFGS1989,byrdLimitedMemoryAlgorithm1995}, a quasi-Newton optimisation method, is applied. L-BFGS is particularly effective in fine-tuning parameters, achieving faster convergence near local optima.
	
	\subsubsection{Optimisation of the Network's architecture: multigrid strategy}
	\label{sec:multi_level}
	The multigrid training proposed for the FENNI framework in Part I is adapted for the tensor decomposition. multigrid training can be set up based on the interpretability of the FENNI networks in which the tensor decomposition is made. The training starts with a very coarse space discretisation to pre-train the tensor decomposition (the idea of starting with coarse modes can be traced back to \parencite{giacomaOptimalPrioriReduced2015}). Once the coarse convergence is reached, a refinement step is performed to get a finer mesh. The associated model is initialised by the previously trained coarse one before being trained. The training of the finer model requires much fewer iterations as most of the training has already been performed at a low cost on the coarse model. Let $\left\{ \overline{\vect{u}}_i^{c} \right\}_{i \in \llbracket 1,m\rrbracket}$ be the coarse modes. Then the finer model can be initialised by setting the nodal values of its finer modes $\left\{ \overline{\vect{u}}_i^{f} \right\}_{i \in \llbracket 1,m\rrbracket}$ as the result of the evaluation of the coarse model at the new fine nodal coordinates. Thus, the fine space mode nodal values read
	\begin{equation}
		\overline{u}_{\ell, i }^{f} = \overline{\vect{u}}_i^c\left(\vect{x}_{\ell}\right), \quad \forall \ell \in \llbracket 1,N_{f}\rrbracket, \forall i \in \llbracket 1,m\rrbracket.
	\end{equation}
	To prevent unnecessary growth of the ROB during multigrid training, the last mode, which does not contribute to the interpolation of the structure, is removed at each level. This ensures the model's efficiency by discarding modes that are not useful by design.
	
	It can be noted that such a training strategy merges the convergence study regarding the mesh size with the proper training of the FENNI-PGD, allowing to get a solution with the coarsest mesh sufficient to obtain a converged solution. 
	
	\newpage
	\section{Numerical results}
	\label{sec:Num}
	The numerical implementation \parencite{daby-seesaramNeuROM2024b} of the method described in this paper is \href{https://github.com/AlexandreDabySeesaram/NeuROM}{openly available online} \footnote{\url{https://github.com/AlexandreDabySeesaram/NeuROM}}. The results presented in this section can be fully reproduced using the provided interactive demos \parencite{daby-seesaramFENNIIIPGDPaperDemo2024}, which include \href{https://alexandredabyseesaram.github.io/FENNI-II-PGD-paper-demo/}{Jupyter notebooks that can be run directly online}\footnote{\url{https://alexandredabyseesaram.github.io/FENNI-II-PGD-paper-demo/}}.

	\subsection{Examples}
	
	Two examples are used to showcase the method presented in this paper. A 1D bar subjected to a volume force and a 2D square with holes subjected to gravity. 
	
	\subsubsection{1D - bar}
\begin{figure}[hbtp!]
		\savebox{\largestimage}{\resizebox{0.48\linewidth}{!}{\input{Body_force_1D.tex}}}%
		\begin{subfigure}[b]{0.48\linewidth}
			\centering
			\raisebox{\dimexpr.5\ht\largestimage-.5\height}{%
				\includegraphics[width=0.9\linewidth]{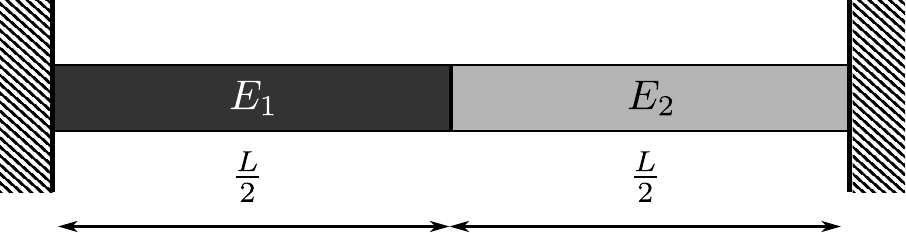}}
			\caption{1D bi-stiffness bar}
			\label{fig:1D_param}
		\end{subfigure}
		\hfill
		\begin{subfigure}[b]{0.48\linewidth}
			\centering
			\usebox{\largestimage}
			\caption{Body force - derived from \parencite{zhangHierarchicalDeeplearningNeural2021}}
			\label{fig:body_f_1D}
		\end{subfigure}
		\caption{1D reference problem}
		\label{fig:ref_1D}
	\end{figure}

	The 1D example is derived from the one used in \parencite{zhangHierarchicalDeeplearningNeural2021}, which has been modified into a parametrised problem to illustrate the proposed FENNI-PGD.
	The 1D bar (of section area $A = \SI{1}{mm\squared}$) highlights two different stiffness parametrised by two young moduli $E_1$ and $E_2$. The first half $\left[ 0,L/2 \right]$ of the bar of length $L = \SI{10}{mm}$ has a stiffness $E_1$ while the second half $\left[ L/1,L \right]$ has a stiffness $E_2$ as illustrated in \cref{fig:1D_param}. The bar is subjected to the discretised volume force 
	
	\begin{equation}
		f\left(x\right) = -\frac{\left(4 \pi^2 \left(x - x_1\right)^2 - 2 \pi\right)}{e^{\left(\pi (x - x_1)^2\right)}} - \frac{\left(8 \pi^2 \left(x - x_2\right)^2 - 4 \pi\right)}{e^{\left(\pi \left(x - x_2\right)^2\right)}},
	\end{equation}
	plotted in \cref{fig:body_f_1D}.
	Note that a simplified example with a single parameter can be obtained by constraining $E_1 = E_2 = E$.

	\subsubsection{2D - square with holes}
	\label{sec:2D_pb_setting}
	
	The second example consists in a square geometry with three holes of different radii. The top and bottom edges are submitted to zero Dirichlet boundary conditions, while the left and right edges are free. The dimensions of the plate are given in \cref{table:Dimensions2DPlate}.
	The plate is immersed in a gravity field at a varying angle $\theta$.

	\begin{table}[hbtp!]
		\label{table:Dimensions2DPlate}
		\centering
		\begin{tabular}{cccccccc}\toprule
			\multicolumn{5}{c}{Lengths} & \multicolumn{3}{c}{Diameters}
			\\\cmidrule(lr){1-5}\cmidrule(lr){6-8}
			$L$ & $L_1$ & $L_2$  & $L_3$ & $L_4$  & $d_1$  & $d_2$ & $d_3$   \\\hline \\[1pt]
			$\SI{10}{mm}$ & $\frac{L}{4}$ & $\frac{3L}{4}$  & $\frac{L}{4}$ & $\frac{3L}{4}$  & $\frac{L}{3}$  & $\frac{L}{5}$ & $\frac{L}{10}$   \\\bottomrule
		\end{tabular}
				\caption{Dimensions of the 2D plate}
	\end{table}
	
	\begin{figure}[H]
		\centering
		\includegraphics[width = 0.7\linewidth]{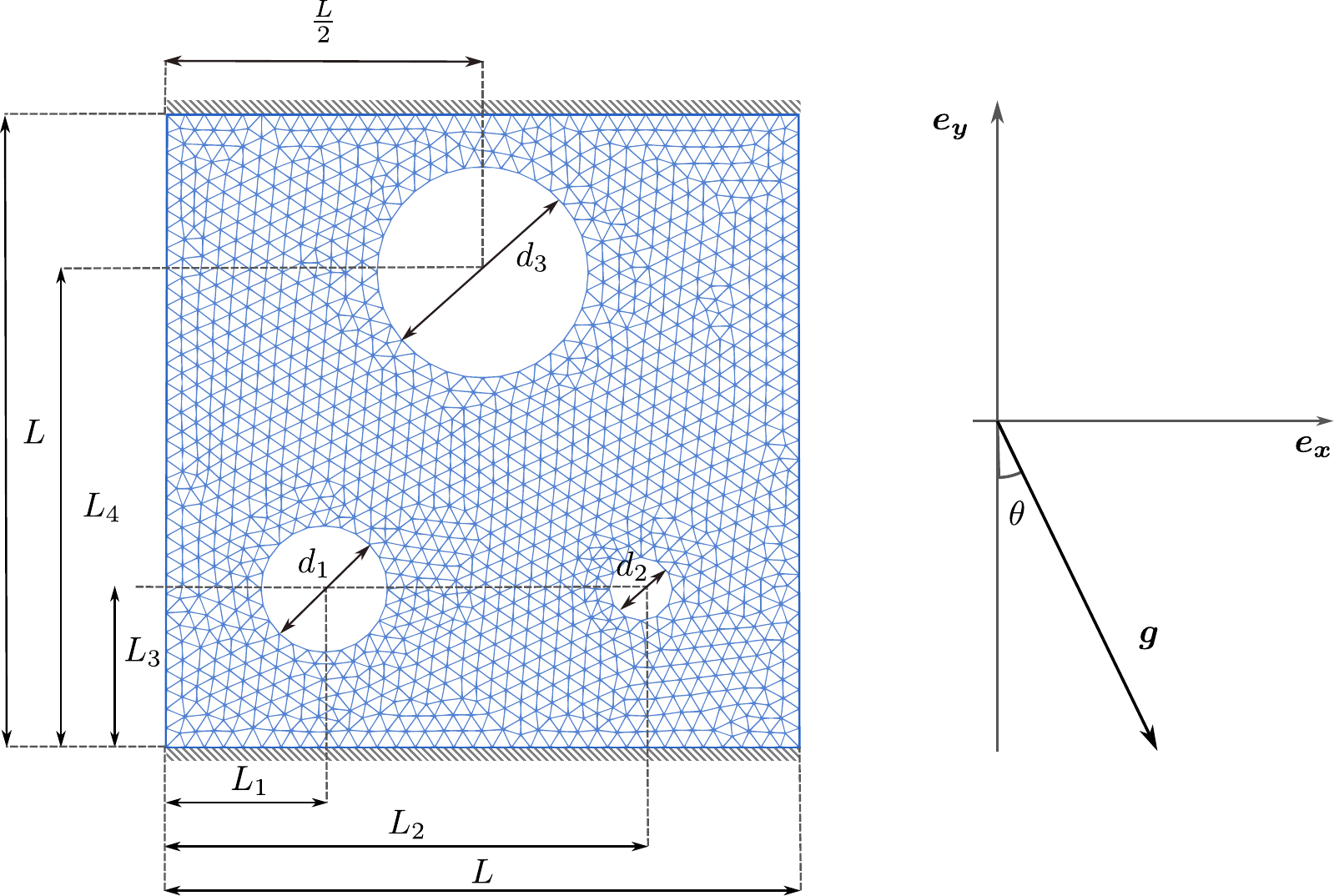}
		\caption{2D reference problem}
		\label{fig:ref_2D}
	\end{figure}

	\subsection{Effectiveness of the tensor decomposition in the FENNI-PGD framework}
	
	This section aims to show the behaviour of the method on a simple 1D example with one or two extra parameters. It investigates the efficiency of the tensor decomposition produced by the method and compares two ways of getting this tensor decomposition: a sequential manner in which the modes are trained one after the other and a simultaneous manner where the modes are still added in a greedy way but where all the modes of the tensor decomposition are trained.

	\subsubsection{Illustration of minimal tensor decomposition}
	
	First, a 1D toy example with an analytical solution is used to show that in a fully separable framework, the method retrieves the minimal tensor decomposition consisting of a single pair of modes. 
	
	\begin{figure}[hbtp!]
		\centering
		\begin{subfigure}[t]{0.48\linewidth}
			\centering
			\resizebox{\linewidth}{!}{\input{L2_1D_Mono}}
			\caption{Evolution of the relative error and size of the reduced-order basis}
			\label{fig:L2_ROB_1D_Mono_Monolevel}
		\end{subfigure}
		\hfill
		\begin{subfigure}[t]{0.48\linewidth}
			\centering
			\resizebox{\linewidth}{!}{\input{Loss_1D_Mono}}
			\caption{Evolution of the loss and size of the reduced-order basis}
			\label{fig:Loss_ROB_1D_Mono_Monolevel}
		\end{subfigure}
		\begin{subfigure}[t]{0.48\linewidth}
			\centering
			\resizebox{\linewidth}{!}{\input{LossDecay_1D_Mono}}
			\caption{Rate of loss decay and reduced-order basis size's evolution}
			\label{fig:d_Loss_ROB_1D_Mono_Monolevel}
		\end{subfigure}
		\caption{Mono-parameter 1D problem - Investigation of the convergence of the reduced-order model and of the evolution of the size of the reduced-order basis. $N =30$.}
		\label{fig:1D_MonoPara_convergence_MonoLevel}
	\end{figure}
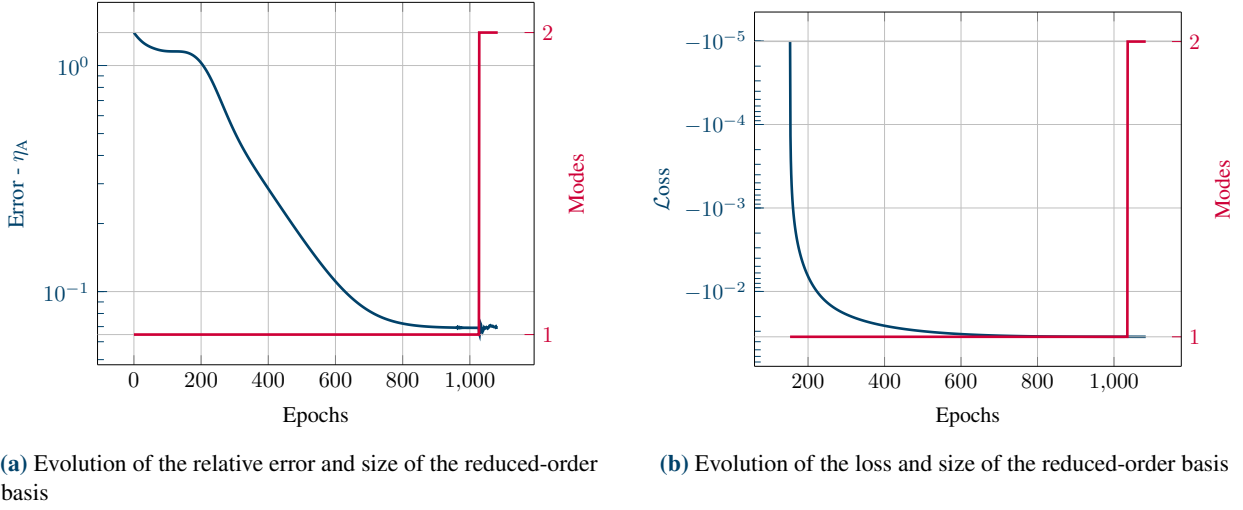
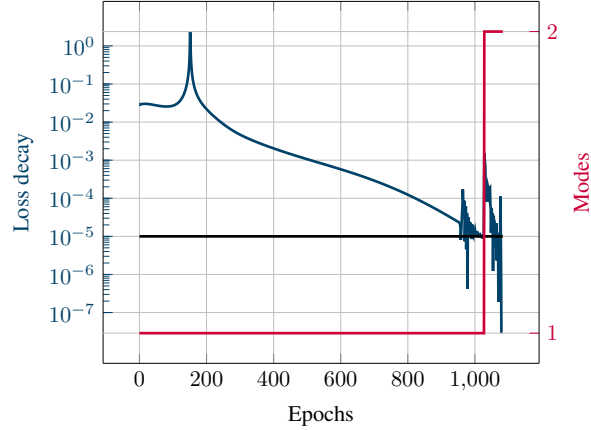

	This example provides an initial illustration of the proposed training strategy. We analyse the evolution of the $L_2$ error throughout the training process and compare it with the decay of training loss to gain insights into the model's performance and convergence.
	The decay of the $L_2$ error is evaluated as
	\begin{equation}
		\eta = \frac{\Vert\overline{ \vect{u}_{NN}-\vect{u}_{\text{exact}}}\Vert}{\Vert\overline{ \vect{u}_{\text{exact}}}\Vert},
	\end{equation}
	with $\overline{\square}$ the average of the quantity $\square$ over the parametric samples of $E$. The decay of the error follows the decrease of the loss $\mathcal{L}$ expressed as the averaged potential energy over the parametric space as shown by \cref{fig:L2_ROB_1D_Mono_Monolevel} and \cref{fig:Loss_ROB_1D_Mono_Monolevel}.
	\cref{fig:1D_MonoPara_convergence_MonoLevel} illustrates the training behaviour of the 1D example on a $30$ nodes mesh: It shows that the greedy training strategy retrieves the single-mode structure required for a correct interpolation of the solution and matches the analytical solution. The analytical solution being
	\begin{equation}
		\begin{split}
			\vect{u} = \frac{1}{AE} \left(\exp\left(- \pi \left( x-x_1 \right)^2 \right) - \exp\left(- \pi x_1^2\right) \right) &+ \frac{2}{AE} \left(\exp\left(- \pi \left( x-x_2 \right)^2 \right)  - \exp\left(- \pi x_2^2\right) \right) \\ 
			&- \frac{\exp\left(- \pi x_1^2\right) -\exp\left(- \pi x_2^2\right) }{10AE}x, 
		\end{split}
	\end{equation}
	it can indeed be written using exactly one mode as
	\begin{equation}
		\vect{u}\left(\vect{x},E\right) = \vect{\overline{u}}\left(\vect{x}\right)\lambda\left(E\right).
	\end{equation}

	\subsubsection{Comparison of sequential and simultaneous mode training approaches}
	\label{sec:seq_vs_parallel}
	
	In this section, the bi-parameter version of the 1D example, which requires multiple modes, is used to compare two training approaches: learning the modes sequentially versus learning all the modes simultaneously.

	\begin{figure}[H]
		\centering
		\begin{subfigure}[t]{0.48\linewidth}
			\centering
			\resizebox{\linewidth}{!}{\input{Loss_1D_Bi_seq}}
			\caption{Evolution of the loss and size of the reduced-order basis - sequential strategy}
			\label{fig:Loss_ROB_1D_Bi_Monolevel_sequential}
		\end{subfigure}
		\hfill
		\begin{subfigure}[t]{0.48\linewidth}
			\centering
			\resizebox{\linewidth}{!}{\input{Loss_1D_Bi_seq_zoom}}
			\caption{Evolution of the loss - zoom on the last epochs of training - sequential strategy}
			\label{fig:Loss_ROB_1D_Bi_Monolevel_zoom_sequential}
		\end{subfigure}
		\begin{subfigure}[t]{0.48\linewidth}
			\centering
			\resizebox{\linewidth}{!}{\input{LossDecay_1D_Bi_seq}}
			\caption{Rate of loss decay and reduced-order basis size's evolution - sequential strategy}
			\label{fig:d_Loss_ROB_1D_Bi_Monolevel_sequential}
		\end{subfigure}
		\caption{Bi-parameter 1D problem - Investigation of the convergence of the reduced-order model and of the evolution of the size of the reduced-order basis. $N=70$ - sequential strategy.}
		\label{fig:1D_BiPara_convergence_MonoLevel_sequential}
	\end{figure}
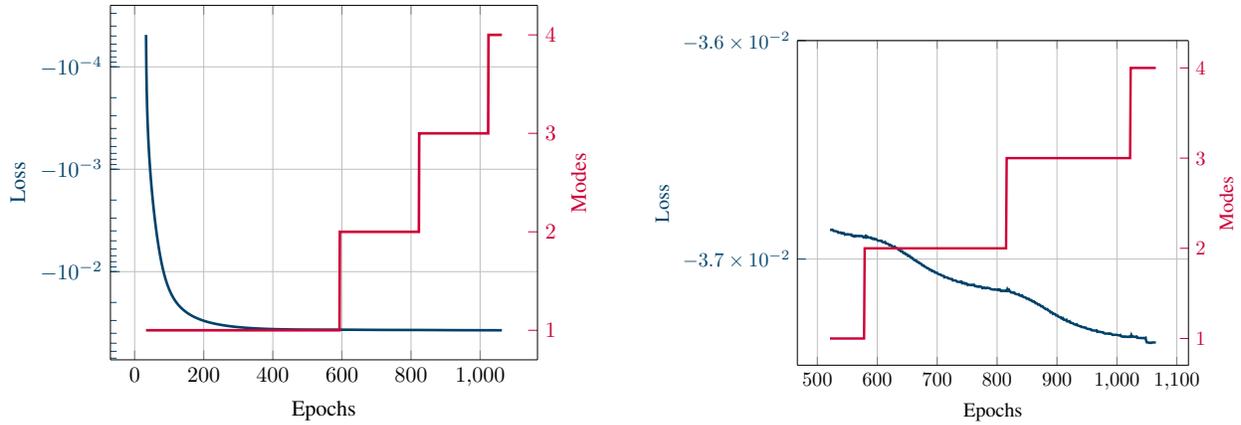
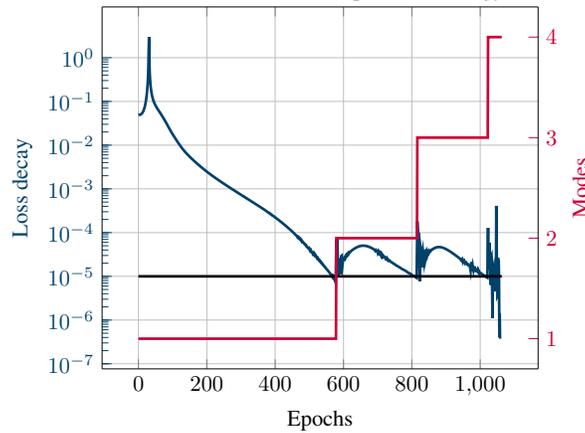

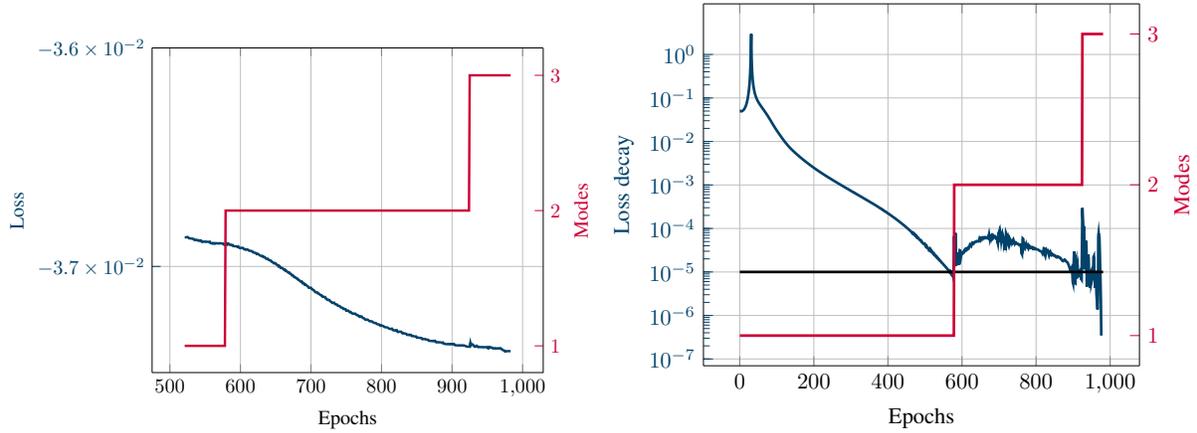
\begin{figure}[H]
		\centering

		\begin{subfigure}[t]{0.48\linewidth}
			\centering
			\resizebox{\linewidth}{!}{\input{Loss_1D_Bi_zoom}}
			\caption{Evolution of the loss - zoom on the last epochs of training - simultaneous training}
			\label{fig:Loss_ROB_1D_Bi_Monolevel_zoom}
		\end{subfigure}
		\begin{subfigure}[t]{0.48\linewidth}
			\centering
			\resizebox{\linewidth}{!}{\input{LossDecay_1D_Bi}}
			\caption{Rate of loss decay and reduced-order basis size's evolution - simultaneous training}
			\label{fig:d_Loss_ROB_1D_Bi_Monolevel}
		\end{subfigure}
		\caption{Bi-parameter 1D problem - Investigation of the convergence of the reduced-order model and of the evolution of the size of the reduced-order basis. $N=70$.}
		\label{fig:1D_BiPara_convergence_MonoLevel}
	\end{figure}

	In the sequential strategy, illustrated in \cref{fig:1D_BiPara_convergence_MonoLevel_sequential}, only the $m$-th mode is trained while the $m-1$ previous modes are frozen. This is closer to what is done in the classical PGD solvers that solve for the reduced-order basis in a greedy manner, sometimes updating the associated parameter modes but keeping space modes fixed. On the other hand, in the simultaneous strategy, illustrated in \cref{fig:1D_BiPara_convergence_MonoLevel}, all $m$ modes are trained at once.

	The comparison of \cref{fig:1D_BiPara_convergence_MonoLevel_sequential} and \cref{fig:1D_BiPara_convergence_MonoLevel} shows that simultaneous training reaches the same loss value as the iterative strategy but with fewer modes.
	\cref{fig:Loss_ROB_1D_Bi_Monolevel_zoom} and \cref{fig:d_Loss_ROB_1D_Bi_Monolevel} shows how adding a mode allows the decay of the loss to speed up. In a similar manner as in the mono-parameter case, the stagnation is reached after adding one last mode that, by construction, does not improve the tensor decomposition of the solution. The simultaneous $m$ modes training will, therefore, be employed in the remainder of the paper.

	\subsubsection{Quantification of the error distribution of the surrogate model solutions}

	\begin{figure}[hbtp!]
		\begin{minipage}[t][][t]{.32\linewidth}
			\centering
			\begin{subfigure}[t]{\linewidth}
				\centering
				\resizebox{\linewidth}{!}{\input{Loss_2D_Bi}}
				\caption{Evolution of the loss and size of the reduced-order basis}
				\label{fig:Loss_ROB_2D}
			\end{subfigure}
			\begin{subfigure}[t]{\linewidth}
				\centering
				\resizebox{\linewidth}{!}{\input{Loss_2D_Bi_zoom}}
				\caption{Evolution of the loss - zoom on the last epochs of training}
				\label{fig:Loss_ROB_2D_zoom}
			\end{subfigure}
			
			\begin{subfigure}[t]{\linewidth}
				\centering
				\resizebox{\linewidth}{!}{\input{LossDecay_2D_Bi}}
				\caption{Rate of loss decay and reduced-order basis size's evolution}
				\label{fig:d_Loss_ROB_2D}
			\end{subfigure}
			\caption{Two-parameter 2D problem - Investigation of the convergence of the reduced-order model and of the evolution of the size of the reduced-order basis.}
			\label{fig:conv_BiStiff_2D_Linear}
		\end{minipage}\hfill
		\begin{minipage}[t]{.32\linewidth}
			\centering
			\begin{subfigure}[t]{\linewidth}
				\includegraphics[width=0.9\linewidth]{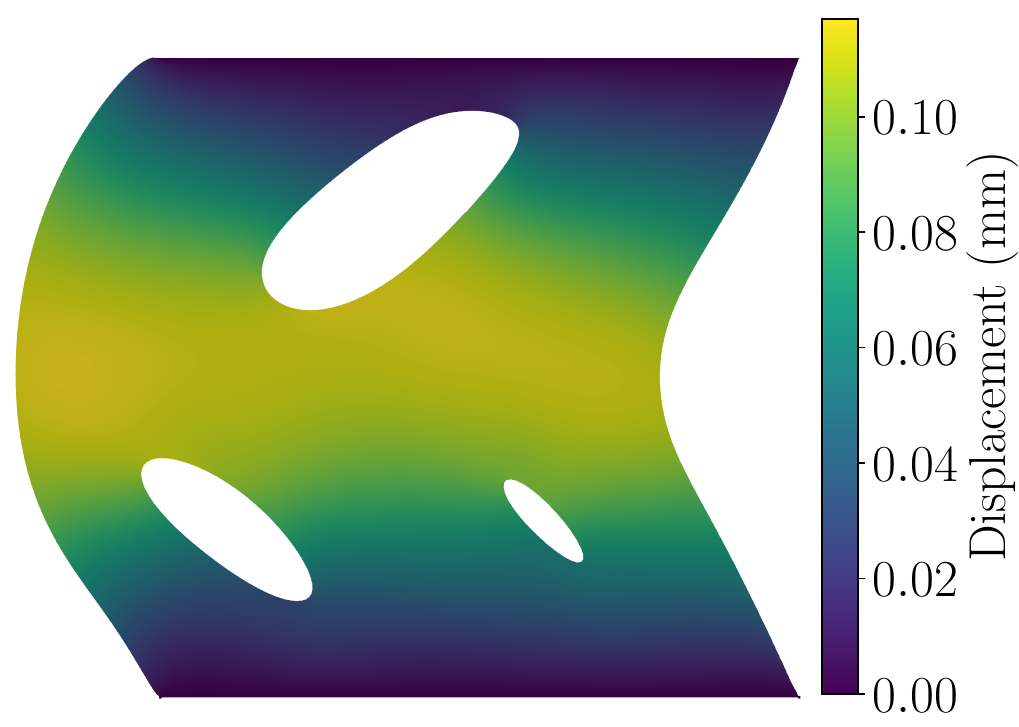}
				\caption{Configuration~1: \\ $\left(E = \SI{3.8e-3}{MPa}, ~\theta = \SI{241}{\degree}\right)$ }
				\label{fig:2D_ROM_p1}
			\end{subfigure}
			\hfil
			\begin{subfigure}[t]{\linewidth}
				\includegraphics[width=0.9\linewidth]{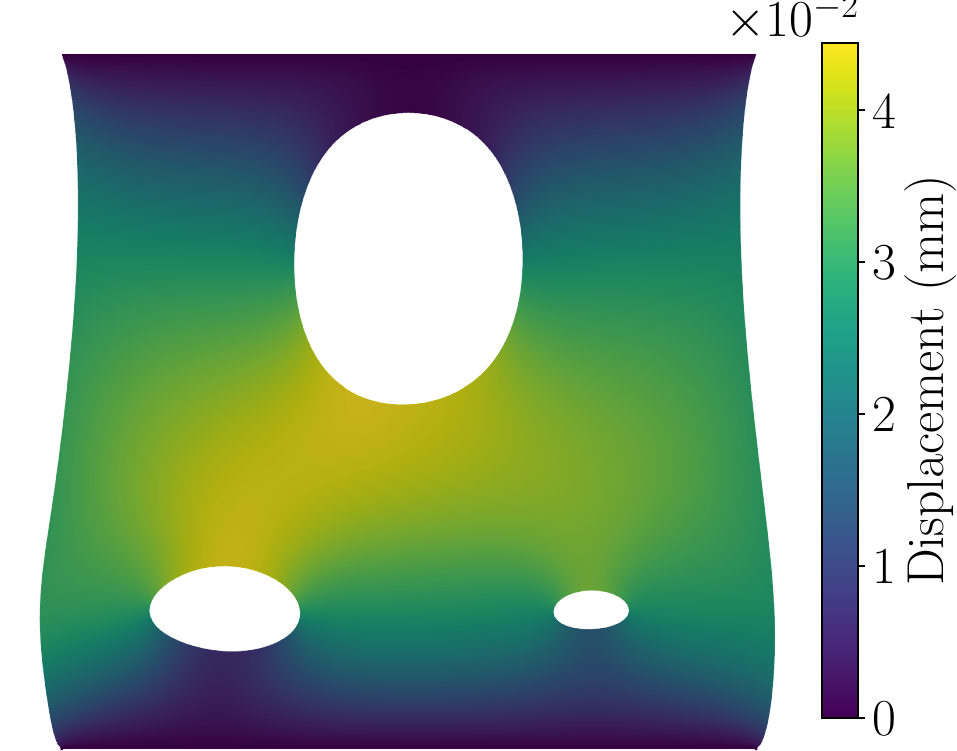}
				\caption{Configuration~2: \\ $\left(E = \SI{3.1e-3}{MPa}, ~\theta = \SI{0}{\degree}\right)$ }
				\label{fig:2D_ROM_p2}
			\end{subfigure}
			
			\begin{subfigure}[t]{\linewidth}
				\includegraphics[width=0.9\linewidth]{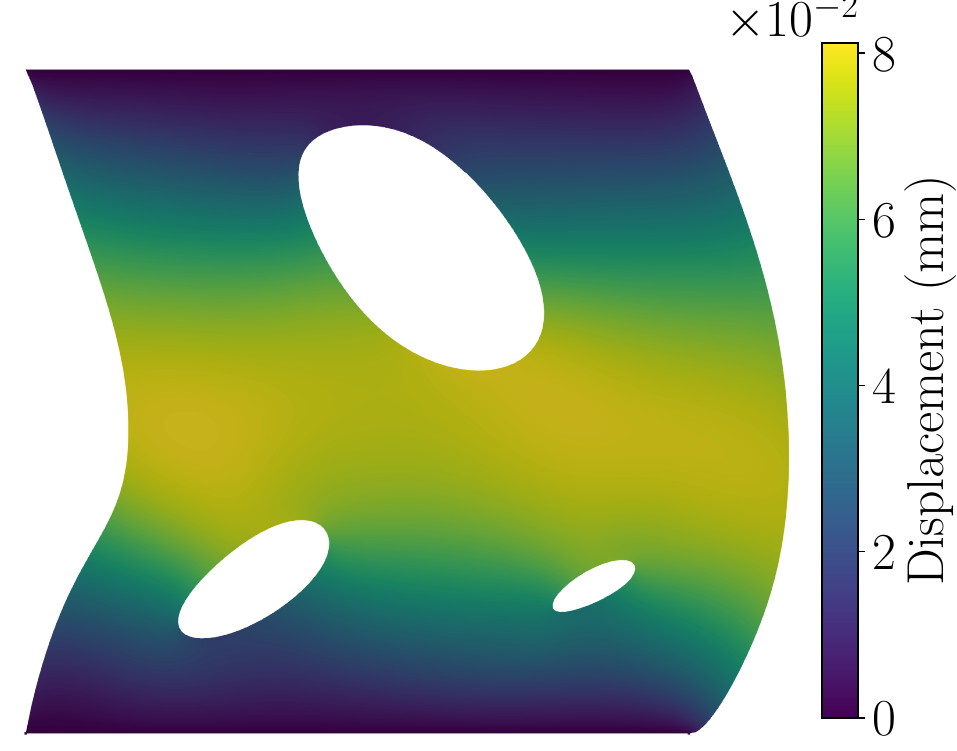}
				\caption{Configuration~3: \\ $\left(E = \SI{4.6e-3}{MPa}, ~\theta = \SI{46}{\degree}\right)$ }
				\label{fig:2D_ROM_p3}
			\end{subfigure}
			\caption{Reduced-order model evaluated at three parameter configurations - norm of displacement plotted on the deformed structure (scaling factor of $20$).}
			\label{fig:2D_ROM_3_config}
		\end{minipage}\hfill
		\begin{minipage}[t][][t]{.32\linewidth}
			\centering
			\begin{subfigure}[t]{\linewidth}
				\centering
				\includegraphics[width=0.95\linewidth]{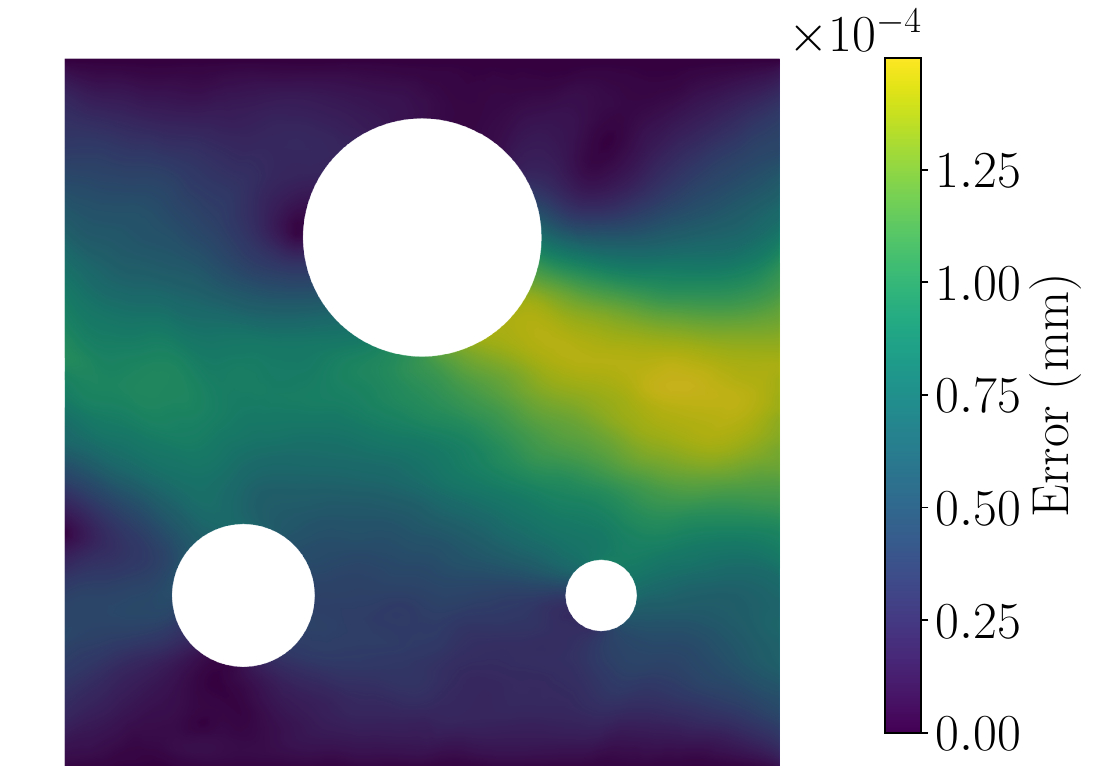}
				
				\caption{Configuration~1: \\ $\left(E = \SI{3.8e-3}{MPa}, ~\theta = \SI{241}{\degree}\right)$ }
				
			\end{subfigure}
			\begin{subfigure}[t]{\linewidth}
				\centering
				\includegraphics[width=0.95\linewidth]{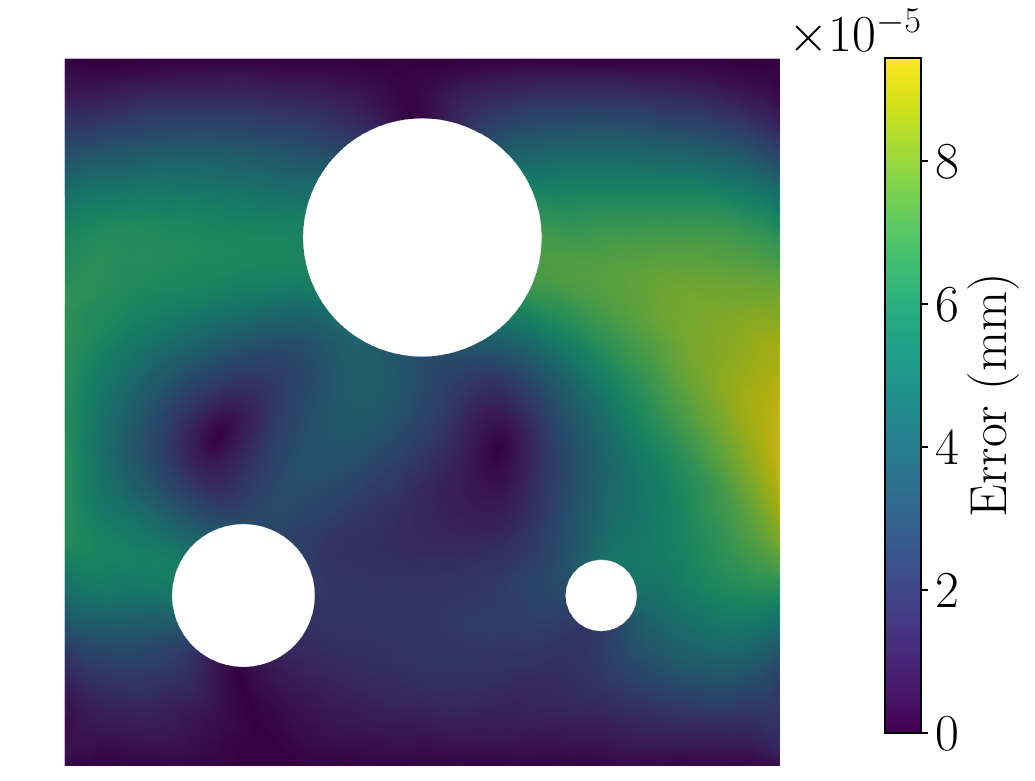}
				\caption{Configuration~2: \\ $\left(E = \SI{3.1e-3}{MPa}, ~\theta = \SI{0}{\degree}\right)$ }
				
			\end{subfigure}
			
			\begin{subfigure}[t]{\linewidth}
				\centering
				\includegraphics[width=0.95\linewidth]{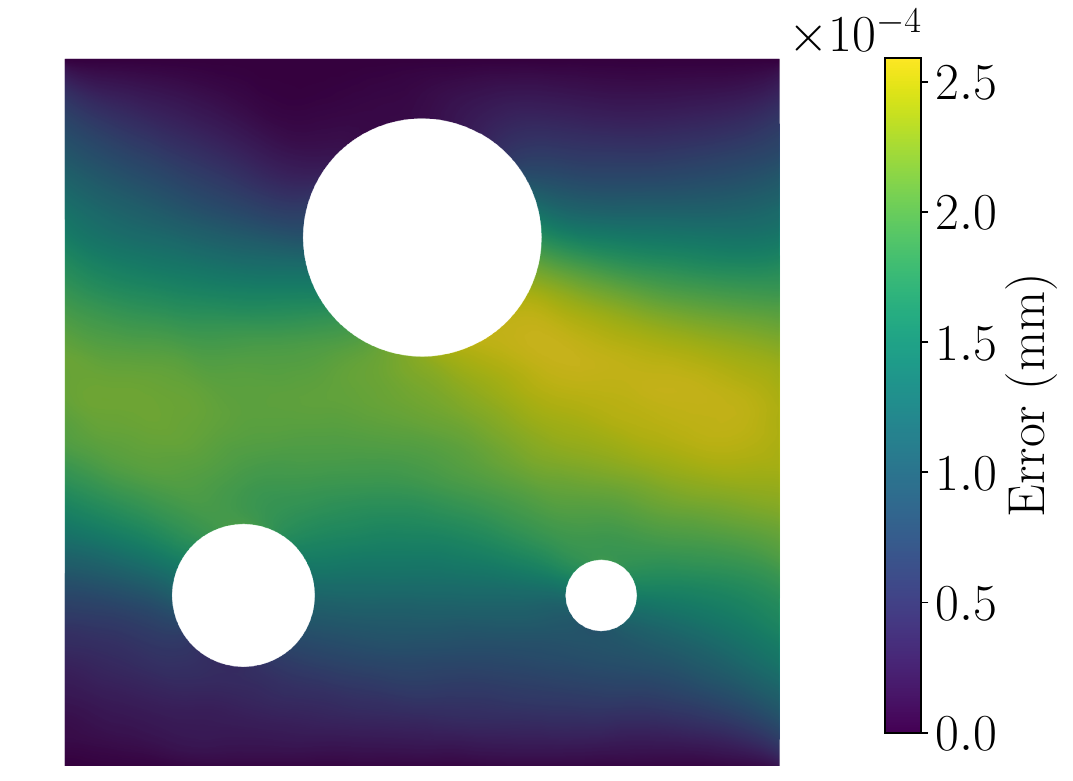}
				\caption{Configuration~3: \\ $\left(E = \SI{4.6e-3}{MPa}, ~\theta = \SI{46}{\degree}\right)$ }
				
			\end{subfigure}
			\caption{Error of the ROM evaluated at three parameter configurations - norm of the error plotted on the deformed structure (scaling factor of $20$).}
			\label{fig:2D_ROM_3_config_error}
		\end{minipage}\hfill
	\end{figure}

	In this section, the accuracy of the 2D surrogate model is evaluated by comparing its output with the solution obtained by the finite element method.
	\cref{fig:Loss_ROB_2D} shows the loss decay and ROB evolution.  Zooming in on the final epochs, \cref{fig:Loss_ROB_2D_zoom} reveals the impact of the addition of the second mode without visual compression artefacts due to the logarithmic scale. Lastly, \cref{fig:d_Loss_ROB_2D} shows the evolution of the stopping criterion and confirms the speed-up of the loss decay following the addition of the second mode while adding the third mode leads to stagnation of the loss thus meaning that the stagnation is not due to the insufficient size of the ROB but rather to the convergence being reached. The algorithm stops shortly after adding the third mode.
	\cref{fig:2D_ROM_3_config} shows the 2D surrogate model evaluated for three distinct parameter configurations.  \cref{table_error_2D} shows the error 
	$$ \eta_{\text{FEM}} =  \frac{\left\Vert \vect{u}_{\text{ROM}}\left(E,
		\theta\right) - \vect{u}_{\text{FEM}}\left(E,
		\theta\right) \right\Vert}{\left\Vert \vect{u}_{\text{FEM}}\left(E,
		\theta\right) \right\Vert}$$
	of the surrogate model compared to the finite element solution. The FEM solution is computed for each parametric configuration while the surrogate model, once trained, is only evaluated for each configuration.

	\begin{table}
		\centering
		\begin{tabular}{|c|c|c|}
			\hline
			$E$ ($\text{kPa}$) & $\theta$ (deg) & Relative error \\
			\hline
			3.80  & 90  & $1.12 \times 10^{-3}$ \\
			3.80  & 241  & $8.72 \times 10^{-4}$ \\
			3.14  & 0     & $1.50 \times 10^{-3}$ \\
			4.09  & 212  & $8.61 \times 10^{-3}$ \\
			4.09  & 179  & $9.32 \times 10^{-3}$ \\
			4.62  & 46  & $2.72 \times 10^{-3}$ \\
			5.01  & 129  & $5.35 \times 10^{-3}$ \\
			6.75  & 312  & $1.23 \times 10^{-3}$ \\
			\hline
		\end{tabular}
		\caption{Error values for each pair of parameters $E$ and $\theta$.}
		\label{table_error_2D}
	\end{table}
	
	The surrogate model exhibits very acceptable levels of errors. Indeed,  \cref{table_error_2D} shows that all eight parametric combinations present a relative error inferior to one percent.  \cref{fig:2D_ROM_3_config_error} shows the error map on the geometry deformed by the error with the same scaling factor of $20$ used in \cref{fig:2D_ROM_3_config}. The error map highlights values three to four orders of magnitude smaller than the solution, again highlighting the good accuracy of the surrogate model.

	\subsection{Optimisation of the Network's architecture: benefits of transfer-learning through interpretability}
	\label{sec:results_ML_training}
	This section aims at showcasing the benefits of using an interpretable interpolation of the different modes in the FENNI. Indeed, the interpretability allows transferring the information from one model to another one by initialising the corresponding parameter of the latter with the interpolated values of the former as described in \cref{sec:multi_level}. The tensor decomposition can, therefore, first be computed on a coarse mesh and then only refined on a finer mesh, thus drastically reducing the cost of finding the fine tensor decomposition. 
	
\begin{figure}[hbtp!]
		\centering
		\includegraphics[width=\linewidth]{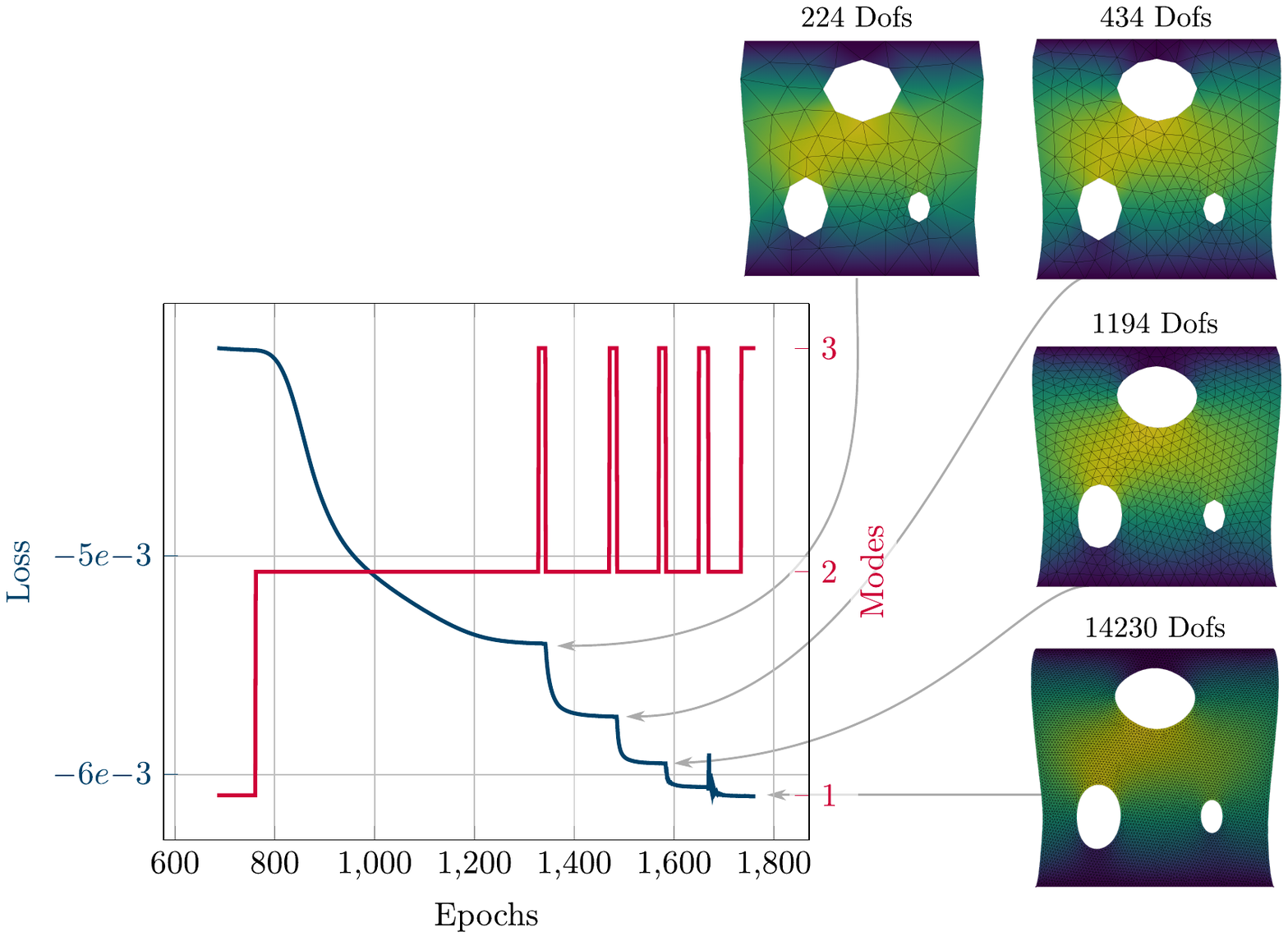}
		\caption{multigrid (5 levels) training convergence for the 2D case and illustration of the refinement of the ROM evaluated for the parametric configuration $\left(E = \SI{5e-3}{MPa}, \theta=\pi \right)$.}
		\label{fig:2D_multi_level_convergence}
	\end{figure}

	The training starts with a coarse mesh, then when convergence is reached, the mesh is refined, and the new model is initialised with the previously trained model. \cref{fig:2D_multi_level_convergence} shows a 5-level training strategy.

	\cref{fig:2D_multi_level_convergence} shows that $80 \%$ of the training is done on the coarse mesh at a low cost while the fine-tuning on the finer meshes only represents the remaining $20 \%$ of the total number of epochs. In such a multigrid approach, the mesh convergence study is performed simultaneously with the training. The modal structure of the ROM is conserved throughout the multigrid training, highlighting the benefits of pre-learning a coarse version of the modes before fine-tuning them.

	\subsection{Non-linear elasticity}
	\label{sec:SVK}
	A non-linear finite strain elastic behaviour has been implemented to describe better soft tissues often modelled as hyperelastic materials to get closer to the target application of an organ digital twin. The framework's versatility is therefore highlighted as the only change required concerns the definition of the loss. 
	
	For the sake of simplicity, the chosen behaviour is the Saint Venant-Kirchhoff law , which is described through the free energy
	
	\begin{equation}
		\Psi := \mu \matrice{E}:\matrice{E} + \frac{\lambda}{2} \mathrm{tr}\left(\matrice{E}\right)^2,
	\end{equation}
	with $\matrice{E} := \frac{1}{2}\left(\matrice{\nabla}\vect{u}^T + \matrice{\nabla}\vect{u} + \matrice{\nabla}\vect{u}^T \matrice{\nabla}\vect{u}\right)$ the Green-Lagrange strain tensor and $\left(\lambda, \mu \right)$ the Lamé coefficients. The Saint Venant-Kirchhoff model is known to be unstable in large compression, but as soft tissues are often described as quasi-incompressible, a Poisson's ratio of $\nu = 0.49$ has been chosen, enforcing the quasi-incompressibility of the elastic behaviour \parencite{letallecNumericalMethodsNonlinear1994b}. 
	
\begin{figure}[H]
		\begin{minipage}[t][][t]{.47\linewidth}
			\centering
			\begin{subfigure}[t]{\linewidth}
				\centering
				\resizebox{\linewidth}{!}{\input{Loss_2D_Bi_SVK}}
				\caption{Evolution of the loss and size of the reduced-order basis}
				\label{fig:Loss_ROB_2D_SVK}
			\end{subfigure}
			\begin{subfigure}[t]{\linewidth}
				\centering
				\resizebox{\linewidth}{!}{\input{Loss_2D_Bi_zoom_SVK}}
				\caption{Evolution of the loss - zoom on the last epochs of training}
				\label{fig:Loss_ROB_2D_zoom_SVK}
			\end{subfigure}
			
			\begin{subfigure}[t]{\linewidth}
				\centering
				\resizebox{\linewidth}{!}{\input{LossDecay_2D_Bi_SVK}}
				\caption{Rate of loss decay and reduced-order basis size's evolution}
				\label{fig:d_Loss_ROB_2D_SVK}
			\end{subfigure}
			
			\caption{Two-parameter 2D problem with SVK behaviour and stiffness-angle parametrisation - investigation of the convergence of the reduced-order model and of the evolution of the size of the reduced-order basis.}
			\label{fig:conv_BiStiff_2D_SVK}
		\end{minipage}\hfill
		\begin{minipage}[t]{.45\linewidth}
			\centering
			\begin{subfigure}[t]{\linewidth}
				\includegraphics[trim={1mm 0 0 0},clip,width=\linewidth]{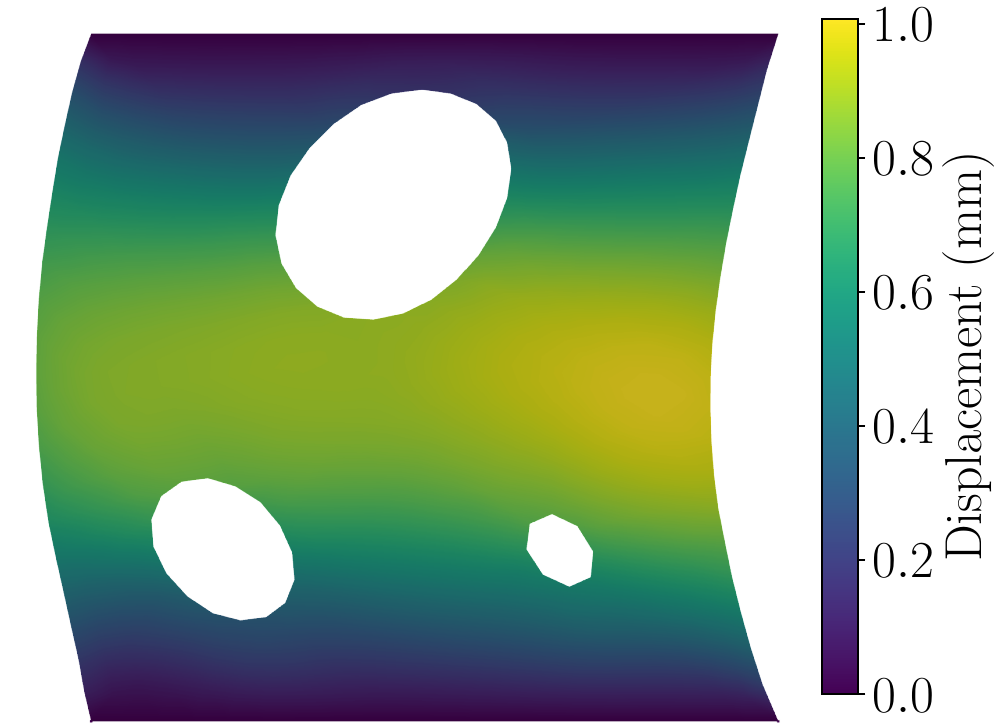}
				\caption{Configuration 1: $\left(E = \SI{3.8e-4}{MPa}, ~\theta = \SI{241}{\degree}\right)$ }
				\label{fig:2D_ROM_p1_SVK}
			\end{subfigure}
			\hfil
			\begin{subfigure}[t]{\linewidth}
				\includegraphics[trim={1mm 0 0 0},clip,width=\linewidth]{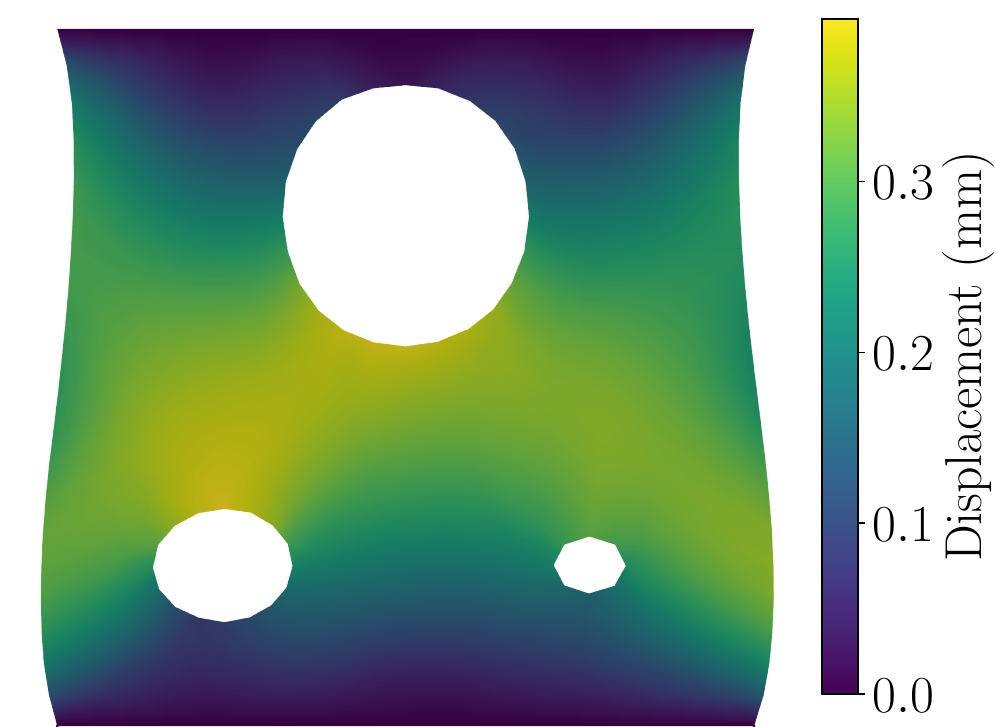}
				\caption{Configuration 2:  $\left(E = \SI{1.1e-4}{MPa}, ~\theta = \SI{0}{\degree}\right)$ }
				\label{fig:2D_ROM_p2_SVK}
			\end{subfigure}
			
			\begin{subfigure}[t]{\linewidth}
				\includegraphics[trim={0 0 0 0},clip,width=\linewidth]{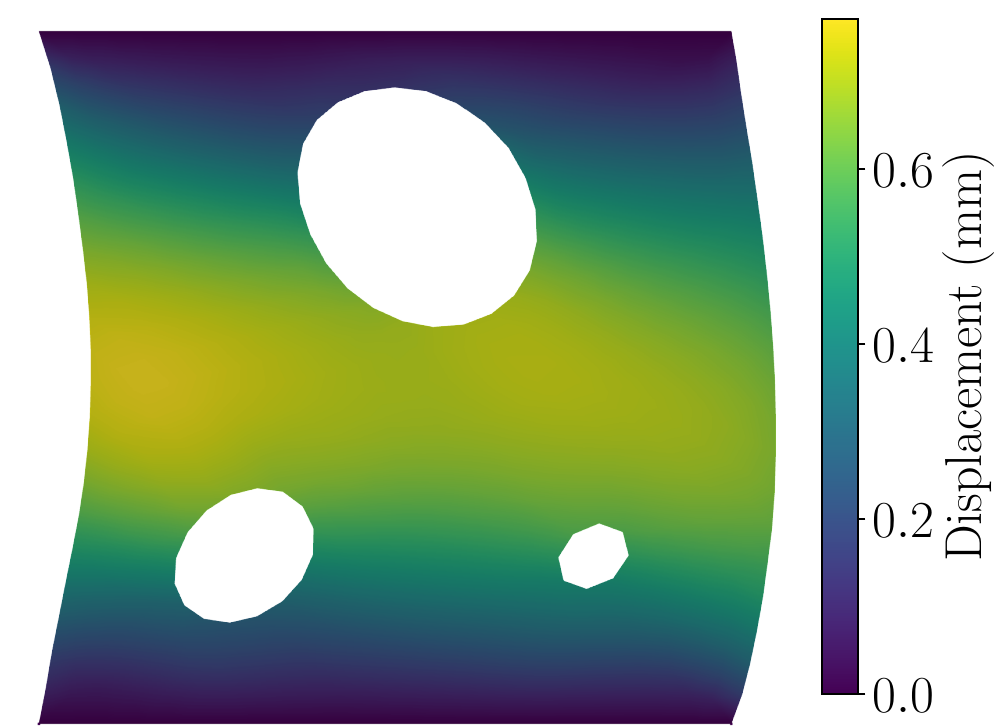}
				\caption{Configuration 3: $\left(E = \SI{4.6e-4}{MPa}, ~\theta = \SI{46}{\degree}\right)$ }
				\label{fig:2D_ROM_p3_SVK}
			\end{subfigure}
			\caption{Reduced-order model evaluated at three parameter configurations - vertical displacement plotted on the deformed structure (scaling factor of $1$) for the SVK behaviour.}
			\label{fig:2D_ROM_3_config_SVK}
		\end{minipage}
	\end{figure}

	The 2D problem presented in \cref{sec:2D_pb_setting} can be solved using this non-linear behaviour by simply changing the definition of the free energy in the definition of the loss. \cref{fig:conv_BiStiff_2D_SVK} shows that the number of epochs required to reach the converged tensor decomposition is similar to what was obtained using the linear loss (see \cref{fig:conv_BiStiff_2D_Linear}) and highlights that on the example as well the greedy training strategy allows to converge towards a low-rank tensor decomposition. 
	
	The surrogate model giving the displacement solution of the parametrised finite strain simulation is evaluated for three parameter configurations, which are shown in \cref{fig:2D_ROM_3_config_SVK}.

	\section{Discussion}
	
	\paragraph{Interpretability} As opposed to PINNs, for which the architecture is fixed during the training, here the interpretability of the framework allows to optimise the architecture of the model by refining the underlying meshes during the training phase, thus removing a major part of the arbitrariness when it comes to choosing the interpolation form for the solution. This feature merges the mesh convergence study with the model's training while also allowing the reuse of a pre-trained model as the initialisation of a (slightly different) new one.  
	
	\paragraph{Solving algorithm}
	
	The framework essentially solves a constrained minimisation problem through gradient-descent-like algorithms facilitated by modern automatic differentiation tools, ensuring efficient results. Relying on deep learning tools to compute the PGD tensor decomposition provides a robust and versatile framework that offers ease of use. Unlike standard (Graph) Neural Network surrogate models, developing a model based on interpretable tensor decomposition presents significant advancements in reducing the computational burden of offline training by enabling transfer learning capabilities. Moreover, it can be noted that the interpretability could also be leveraged to efficiently reuse a trained model as an initial guess for a slightly different surrogate modelling application.

	Concerning the comparison performed in \cref{sec:seq_vs_parallel}, the preferred greedy method depends on the goal of the surrogate model. Indeed, while training the modes simultaneously leads to a more optimal PGD tensor decomposition, it comes at the cost of a more computationally expensive offline training phase. The choice between sequential or simultaneous training, therefore, depends on the priority between a fast evaluation of the model that relies on the smallest decomposition possible or the training efficiency that would be reduced in a setting where fewer parameters are being optimised at once.
	
	\paragraph{Overcoming the curse of dimensionality}
	
	As mentioned in \cref{ap:loss}, the non-linearity involved in the description of the problem might lead to losses that require building the full-order parametrised solution. In such cases where the loss cannot be computed without assembling the different modes into the full solution, the curse of dimensionality leads to exponentially increasing computation costs despite the reasonable number of parameters. This paper only focuses on losses that can be written without building the full solution, namely the linear elasticity and the Saint Venant-Kirchhoff elasticity. Different training strategies should be envisaged for physics that do not fall in this scope. 
	
	Despite this limitation, several other problems still fall within the scope of the proposed method. Indeed, Incompressible Neo-Hookean material, described through the strain energy density
	\begin{equation}
		\begin{cases}
			\Psi = C_1 \left(\mathrm{tr}\left(\matrice{C}-\matrice{1}\right)\right) \\
			\mathrm{det}\left(\matrice{F}\right) = 1,
		\end{cases}
	\end{equation}
	with $\matrice{C}$ the right Cauchy-Green deformation tensor and $\matrice{F}$ is the deformation gradient can be written based on the individual tensors of the tensor decomposition using Einstein's notation in a similar manner as for the Saint Venant-Kirchhoff behaviour.  Note that coupling physics that individually falls within this limitation also lead to a problem that can be straightforwardly tackled with the proposed framework. Strongly coupled poro-elastic or thermo-elastic problems could thus, for instance, be solved within this framework.
	
	\paragraph{Adaptivity of the meshes}
	
	One promising perspective to this framework is the possibility of using mesh adaptivity naturally embedded in the FENNI framework to adapt the parametric space sampling. This framework could also allow each PGD mode to have a different level of mesh refinement, which could prove valuable for the PGD, where some modes capturing global effects could be coarser than those that capture more localised corrections. A first example of this last point is proposed on the 1D example in \cref{sec:mesh_adapt_1D}.
	
	\paragraph{Application to problems exhibiting a larger Kolmogorv width.} The method has been showcased on linear and non-linear cases exhibiting very shallow Kolmogorov widths but is not limited to such cases. \cref{sec:Larger_Kolmogorov}, for instance, succinctly shows the method being applied to a heterogeneous parametrisation of the structure's stiffness, which parametrised solution requires a larger number of modes. Further application of the methodology will be focused on medically relevant cases, highlighting a larger Kolmogorov width to put the method to the test further. In such cases, if the Kolmogorov width depends on the fineness of the mesh, it is expected that more modes associated with smaller wavelength corrections will be added during the multigrid training contrary to what was highlighted on the varying gravity angle's example where the number of required modes was independent of the coarseness of the mesh (see \cref{sec:results_ML_training}). 
	In this context,  the proposed framework allows further hybridising the PGD with more standard deep-learning approaches. A high-dimensional problem could, for instance, be reduced using an auto-encoder, and the resulting latent space could be used in the PGD decomposition. This would open the way to describing highly non-linear parametric problems using a PGD linear subspace, letting the non-linear auto-encoder encode the non-linearity parametrisation.

	\section{Conclusion}
	
	This paper shows that the FENNI framework can be used as an elementary building block to generate the modes of the PGD solution. Therefore, a surrogate model based on a tensor decomposition can be built automatically on the fly, relying on back-propagation and modern tools developed for deep learning approaches. Taking advantage of those tools, finding the tensor decomposition is straightforward and efficient. The training of the model allows us to find the dimensionality, \emph{i.e.} the number of modes of the PGD while solving the PDEs describing the problem. The proposed framework is versatile, as changing the problem by changing the physics or the parametrisation can be done by updating the loss function accordingly without impacting any other aspect of the method, making it adequate for contexts where digital twins of different natures must coexist. This aspect has been illustrated by training linear elasticity models and a finite strain elastic model. Training all modes simultaneously also allows for finding an efficient tensor decomposition, as shown in the bar example, which later translates into a lighter model being evaluated online. Several parametrisation have been shown. The surrogate model has been compared to a finite element solution and shows small relative errors, validating the working principle of the framework.
	
	Following this proof of concept, more challenging examples exhibiting larger Kolmogorov widths should be considered. A first step could be the joint use of the proposed framework with auto-encoder to provide a surrogate model of apparently very high dimensional problems using a reasonable number of parameters in the tensor decomposition.
	
	\section*{Acknowledgements}
	
	This work was supported by the French National Research Agency (ANR) under the grant MLQ-CT (ANR-23-CE17-0046) and  Bertip EUR (ANR 18EURE0002).

	\appendix
	
	\section{Loss computation}
	\label{ap:loss}
	The numerical implementation of the loss can drastically impact the computational cost of the training stage. Indeed, the naive approach of assembling the full-order solution to compute the loss is much more expensive than building the loss based on the tensor decomposition. 
	
	\subsection{Linear elasticity}
	In linear elasticity, defined by the strain energy density
	\begin{equation}
		\Psi := \frac{\lambda}{2}\text{tr}\left(\eps\right)^ 2 + \mu \eps : \eps,
	\end{equation}
	 the loss, given in \cref{eq:loss} corresponding to the potential energy can be written as 
	
	\begin{equation}
		\mathcal{L} =  {E_p}_{\text{int}} + {E_p}_{\text{ext}}, 
	\end{equation}
	where
	\begin{equation}
		\begin{cases}
			{E_p}_{\text{int}} =   \int_{\mathcal{B}}\frac{1}{2} \intV \eps : \ftensor{C} : \eps \dV\mathrm{d}\beta. \\
			{E_p}_{\text{ext}} =   \int_{\mathcal{B}}  \intSn - \vect{F}\cdot \vect{u} \dS + \intV - \vect{f}\cdot\vect{u}\dV\mathrm{d}\beta.
		\end{cases}
		\label{eq:work}
	\end{equation}
	
	Relying on the tensor decomposition, the discrete version of \cref{eq:work} written with Einstein's notation reads
	\begin{equation}
		\begin{cases}
			\frac{2W_{\text{int}}}{\d \alpha \d E} =  K_{ij} \eps_{j}\left(\uPGD_{em}\right) \eps_{i}\left(\uPGD_{el}\right) \mathrm{det}\left(J\right)_{em} \lambda^{\left(1\right)}_{mp} \lambda^{\left(2\right)}_{lt}  \lambda^{\left(1\right)}_{mp} \lambda^{\left(2\right)}_{lt} E_p \\
			\frac{W_{\text{ext}}}{\d \alpha \d E } =  \uPGD_{iem} \mathrm{det}\left(J\right)_{em} g_{it}   \lambda^{\left(1\right)}_{mp} \lambda^{\left(2\right)}_{mt}
		\end{cases}
		\label{eq:work_einstein}
	\end{equation}
	where $K$ is the elementary stiffness matrix and $\mathrm{det}\left(J\right)$ is the determinant of the each element.
	
	The loss can then be calculated without the need to build the full-order solution $\vect{u}\left(\vect{x}, \para\right) $ by relying on Pytorch \parencite{anselPyTorch2Faster2024} implementation of \code{torch.einsum} which uses broadcasting to efficiently compute the scalar output being the loss.
	
	\subsection{Saint Venant-Kirchhoff}
	
	Writing the loss using Einstein's notation remains straightforward for Saint Venant-Kirchhoff models. Due to their relative simplicity, such models are often used to describe soft tissues. It must be noted that this model, being unstable under important compression, is inherently flawed. However, the quasi-incompressibility property of the modelled soft tissues allows reasonable use of this model under finite strain. 
	
	As mentioned in \cref{sec:SVK}, the strain energy in the SVK model reads
	
	\begin{equation}
		\Psi = \mu \matrice{E}:\matrice{E} + \frac{\lambda}{2} \mathrm{tr}\left(\matrice{E}\right)^2
	\end{equation}
	
	with 
	
	\begin{equation}
		\matrice{E} = \frac{1}{2}\left(\matrice{\nabla}\vect{u}^T + \matrice{\nabla}\vect{u} + \matrice{\nabla}\vect{u}^T \cdot \matrice{\nabla}\vect{u} \right)
	\end{equation}
	being the Green-Lagrange tensor.
	
	The loss then reads
	
	\begin{equation}
		\mathcal{L} =  \frac{\mu}{E} W_{E:E} + \frac{\lambda}{E} W_{\text{trE2}} - W_{\text{ext}}
	\end{equation}
	
	with 
	
	\begin{equation}
		\begin{cases}
			W_{E:E} = \frac{1}{2}{W_{E:E}}_1 + \frac{1}{2}{W_{E:E}}_2 + {W_{E:E}}_3 + \frac{1}{4}{W_{E:E}}_4 \\
			W_{\text{trE2}} = {W_{\text{trE2}}}_1 + {W_{\text{trE2}}}_2 + \frac{1}{4}{W_{\text{trE2}}}_3 
		\end{cases}
	\end{equation}
	and,

	\begin{equation}
		\begin{cases}
			{W_{E:E}}_1 =   \mathrm{det}\left(J\right)_{em},\nabla\left(\uPGD_{em}\right)_{xy},\lambda^{\left(0\right)}_{mp},\lambda^{\left(0\right)}_{mt},\nabla\left(\uPGD_{el}\right)_{yx},\lambda^{\left(0\right)}_{lp},\lambda^{\left(1\right)}_{lt},E_p\\  
			{W_{E:E}}_2 =   \mathrm{det}\left(J\right)_{em},\nabla\left(\uPGD_{em}\right)_{xy},\lambda^{\left(0\right)}_{mp},\lambda^{\left(0\right)}_{mt},\nabla\left(\uPGD_{el}\right)_{xy},\lambda^{\left(0\right)}_{lp},\lambda^{\left(1\right)}_{lt},E_p\\  
			{W_{E:E}}_3 =   \mathrm{det}\left(J\right)_{em},\nabla\left(\uPGD_{em}\right)_{zx},\lambda^{\left(0\right)}_{mp},\lambda^{\left(0\right)}_{mt},\nabla\left(\uPGD_{el}\right)_{zy},\lambda^{\left(0\right)}_{lp},\lambda^{\left(1\right)}_{lt},
			\nabla\left(\uPGD_{en}\right)_{yx},\lambda^{\left(0\right)}_{np},\lambda^{\left(1\right)}_{nt},E_p\\  		
			{W_{E:E}}_4 =   \mathrm{det}\left(J\right)_{em},\nabla\left(\uPGD_{em}\right)_{zx},\lambda^{\left(0\right)}_{mp},\lambda^{\left(0\right)}_{mt},\nabla\left(\uPGD_{el}\right)_{zy},\lambda^{\left(0\right)}_{lp},\lambda^{\left(1\right)}_{lt},
			\nabla\left(\uPGD_{en}\right)_{wy},\lambda^{\left(0\right)}_{np},\lambda^{\left(1\right)}_{nt},
			\nabla\left(\uPGD_{eo}\right)_{wx},\lambda^{\left(0\right)}_{op},\lambda^{\left(1\right)}_{ot},E_p\\  		
			{W_{\text{trE2}}}_1 = \mathrm{det}\left(J\right)_{em},\nabla\left(\uPGD_{em}\right)_{xx},\lambda^{\left(0\right)}_{mp},\lambda^{\left(0\right)}_{mt},\nabla\left(\uPGD_{el}\right)_{yy},\lambda^{\left(0\right)}_{lp},\lambda^{\left(1\right)}_{lt},E_p\\
			{W_{\text{trE2}}}_2 =\mathrm{det}\left(J\right)_{em} ,\nabla\left(\uPGD_{em}\right)_{xx} ,\lambda^{\left(0\right)}_{mp},\lambda^{\left(1\right)}_{mt}, \nabla\left(\uPGD_{el}\right)_{XY},\lambda^{\left(0\right)}_{lp},\lambda^{\left(1\right)}_{lt},\nabla\left(\uPGD_{en}\right)_{XY},\lambda^{\left(0\right)}_{np},\lambda^{\left(1\right)}_{nt},E_p \\
			{W_{\text{trE2}}}_3 =\mathrm{det}\left(J\right)_{em} ,\nabla\left(\uPGD_{em}\right)_{yx} ,\lambda^{\left(0\right)}_{mp},\lambda^{\left(1\right)}_{mt}, \nabla\left(\uPGD_{el}\right)_{yx},\lambda^{\left(0\right)}_{lp},\lambda^{\left(1\right)}_{lt},\nabla\left(\uPGD_{en}\right)_{YX},\lambda^{\left(0\right)}_{np},\lambda^{\left(1\right)}_{nt}
			,\nabla\left(\uPGD_{eo}\right)_{YX},\lambda^{\left(0\right)}_{op},\lambda^{\left(1\right)}_{ot}
			,E_p
		\end{cases}
	\end{equation}
	
	\section{Independent mesh adaptivity for each PGD mode}
	\label{sec:mesh_adapt_1D}
	
\begin{figure}[H]
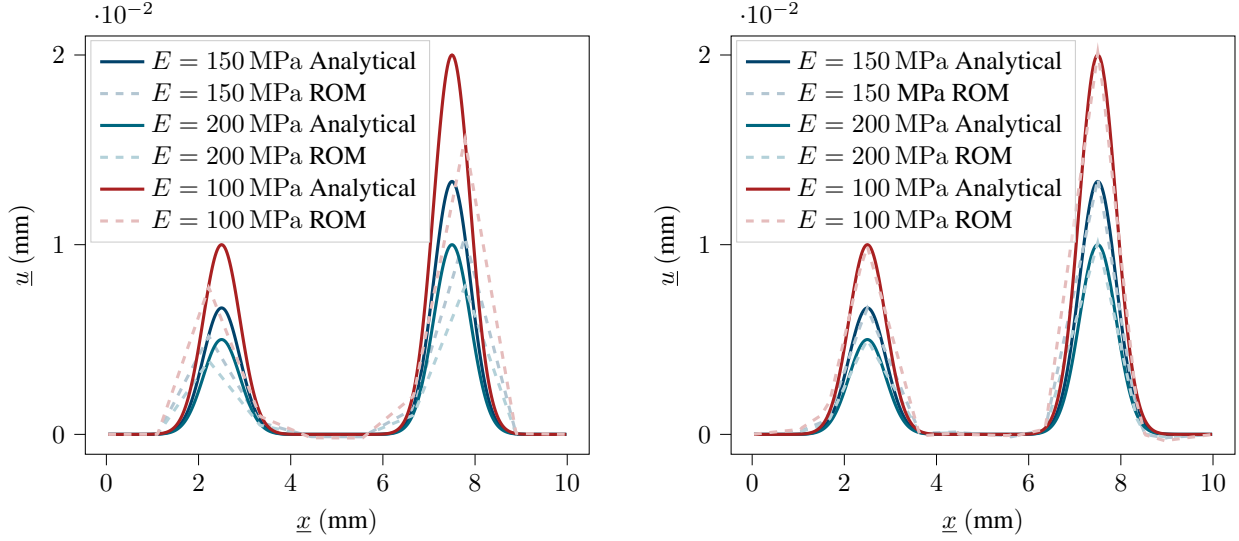

		\begin{subfigure}[t]{0.48\linewidth}
			\centering
			\resizebox{\linewidth}{!}{\input{Para_displacementsPlot_ROM_FOM_10__FrozenMesh__MonoPara1.e-05.tex}}
			\caption{1D mono-stiffness ROM with a frozen mesh - $\eta = 41\%$}
		\end{subfigure}
		\hfill
		\begin{subfigure}[t]{0.48\linewidth}
			\centering
			\resizebox{\linewidth}{!}{\input{Para_displacementsPlot_ROM_FOM_10__FreeMesh__MonoPara1.e-05.tex}}
			\caption{1D mono-stiffness ROM with free meshes - $\eta = 9.5\%$}
		\end{subfigure}
		\caption{Influence of having a trainable mesh on coarse space discretisation, comparison of the ROM with the analytical solution $\left(N = 10 \right)$}
		\label{fig:TrainableMesh_dip_1D}
	\end{figure}

	To illustrate the benefit of letting each mode adapt the underlying mesh, a very coarse mesh of $10$ nodes is used in the 1D example. By allowing r-adaptivity, having different meshes for each mode can lead to an artificially finer mesh by using out-of-phase nodes for the different modes, thus covering a wider space. \cref{fig:TrainableMesh_dip_1D} shows, for instance, that using only $10$ nodes with a fixed mesh leads to a significant interpolation error $41\%$, that can be decreased to $9.5\%$ using still $10$ nodes per mode with free meshes for the modes.

	\section{Inhomogeneous stiffness}
	\label{sec:Larger_Kolmogorov}
	In anticipation of a medical application where two different stiffness zones must be considered, we present a final example featuring a meta-model that enables on-the-fly evaluation of mechanical solutions for a 2D plate subjected to a vertical gravitational field. The plate consists of two regions: 
	
	\begin{equation}
		\begin{cases}
			\text{Left:} x \le \alpha L \\
			\text{Right:} x > \alpha L \\
		\end{cases}
	\end{equation}
	where the left region has a stiffness of $E$ and the right region has a stiffness of $E+\delta E$. This example is illustrated using a linear elasticity behaviour.
	
\begin{figure}[H]
		\begin{minipage}[t][][t]{.47\linewidth}
			\centering
			\begin{subfigure}[t]{\linewidth}
				\centering

				\resizebox{\linewidth}{!}{\input{Loss_2D_Bi_SV}}
				\caption{Evolution of the loss and size of the reduced-order basis}
				\label{fig:Loss_ROB_2D_SV}
			\end{subfigure}

			\begin{subfigure}[t]{\linewidth}
				\centering

				\resizebox{\linewidth}{!}{\input{Loss_2D_Bi_zoom_SV}}
				\caption{Evolution of the loss - zoom on the last epochs of training}
				\label{fig:Loss_ROB_2D_zoom_SV}
			\end{subfigure}
			
			\begin{subfigure}[t]{\linewidth}
				\centering

				\resizebox{\linewidth}{!}{\input{LossDecay_2D_Bi_SV}}
				\caption{Rate of loss decay and reduced-order basis size's evolution}
				\label{fig:d_Loss_ROB_2D_SV}
			\end{subfigure}
			
			\caption{Two-parameter 2D problem with in-homogeneous stiffness- Investigation of the convergence of the reduced-order model and of the evolution of the size of the reduced-order basis.}
			\label{fig:conv_BiStiff_2D}
		\end{minipage}\hfill
		\begin{minipage}[t]{.45\linewidth}
			\centering
			\begin{subfigure}[t]{\linewidth}

				\includegraphics[trim={1mm 0 0 0},clip,width=\linewidth]{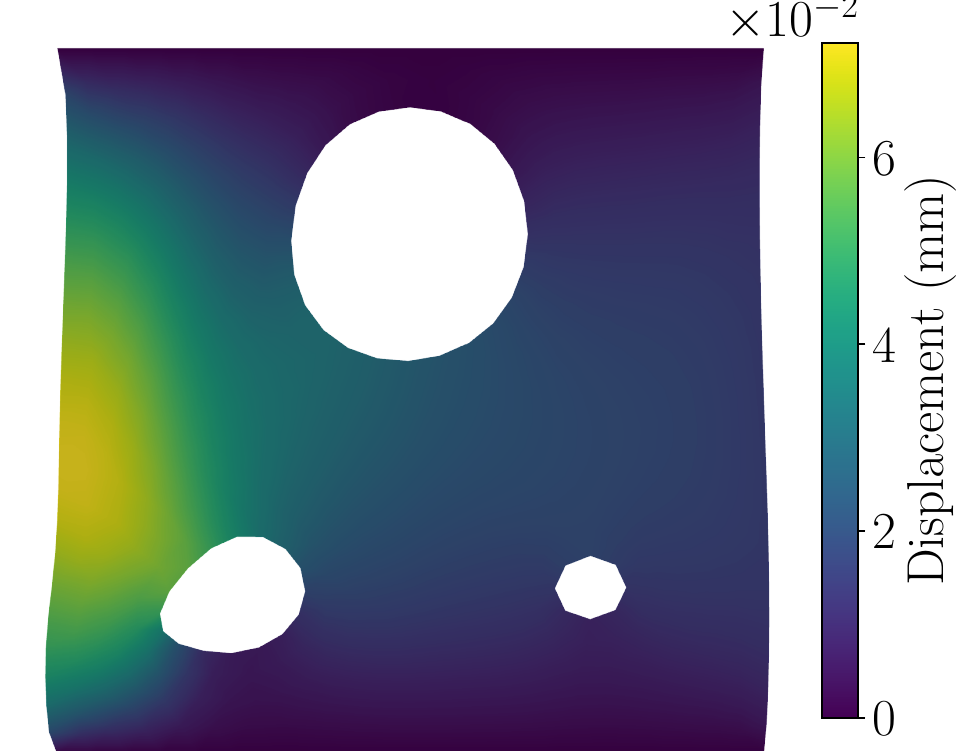}
				\caption{Configuration 1: $\left(\Delta E = \SI{1e-2}{MPa}, ~\alpha = 0.25\right)$ }
				\label{fig:2D_ROM_p1_space_var}
			\end{subfigure}
			\hfil
			\begin{subfigure}[t]{\linewidth}

				\includegraphics[trim={1mm 0 0 0},clip,width=\linewidth]{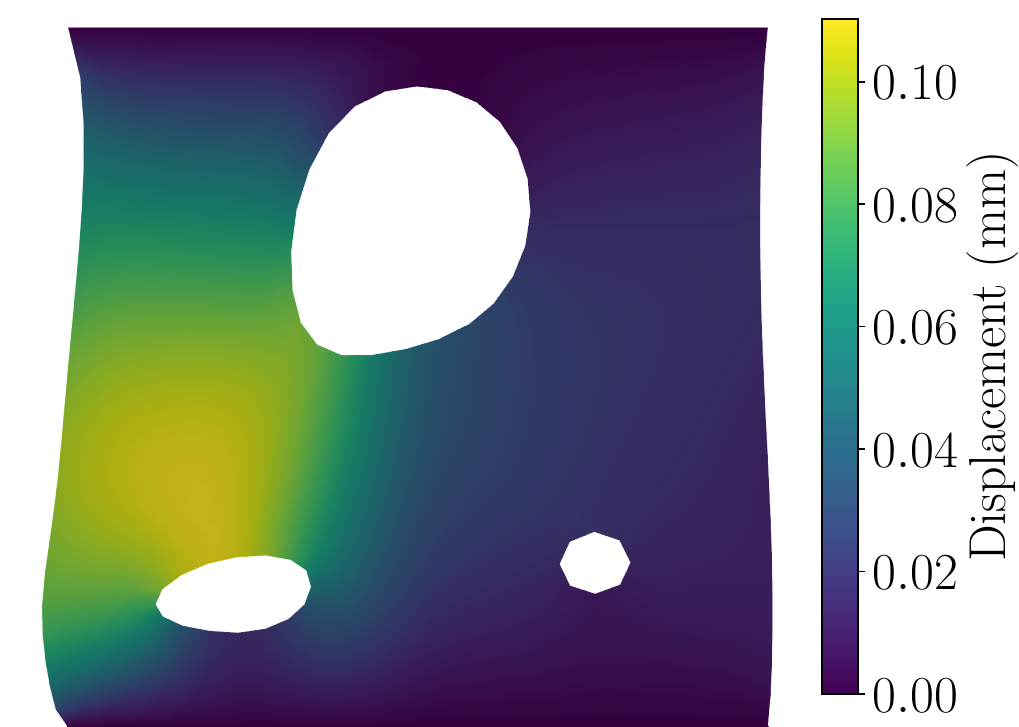}

				\caption{Configuration 2: $\left(\Delta E = \SI{1e-2}{MPa}, ~\alpha = 0.5\right)$ }
				\label{fig:2D_ROM_p2_space_var}
			\end{subfigure}
			
			\begin{subfigure}[t]{\linewidth}

				\includegraphics[trim={0 0 0 0},clip,width=\linewidth]{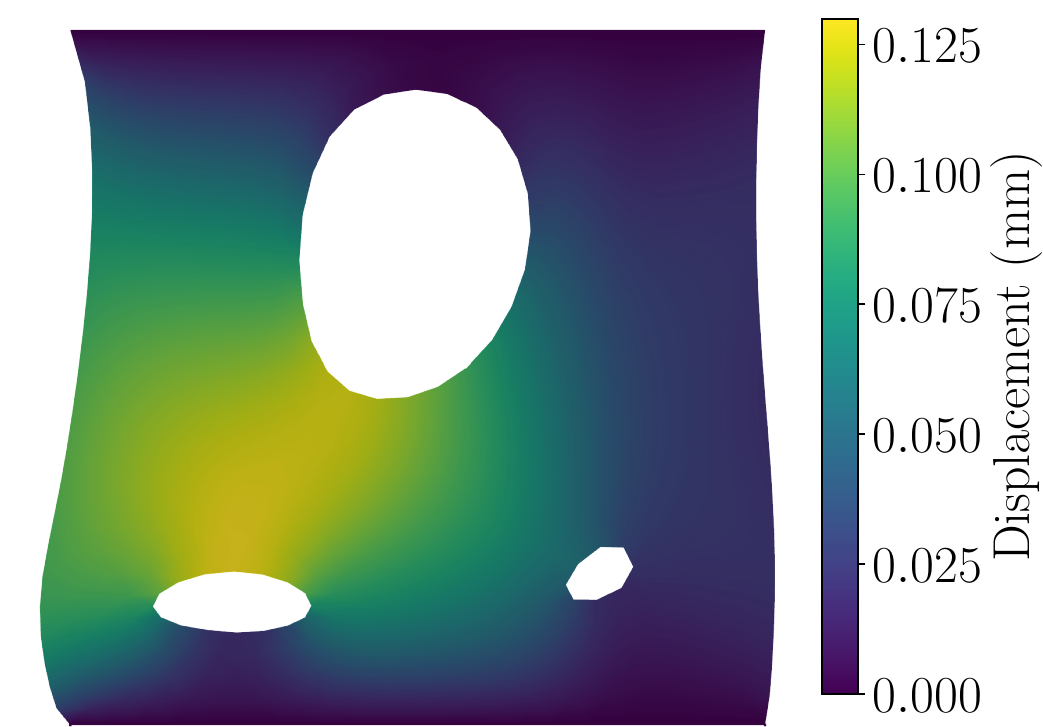}

				\caption{Configuration 3: $\left(\Delta E = \SI{1e-2}{MPa}, ~\alpha = 0.8\right)$ }
				\label{fig:2D_ROM_p3_space_var}
			\end{subfigure}
			\caption{Reduced-order model evaluated at three parameter configurations - vertical displacement plotted on the deformed structure (scaling factor of $10$) with $E_0 = \SI{1e-3}{MPa}$}
			\label{fig:2D_ROM_3_config_space_var}
		\end{minipage}
	\end{figure}

	The meta-model provides a solution that can be evaluated for any given parameters $\left[\delta E, \alpha\right] \in \left[1e-3, 1e-2\right] \bigtimes \left[0,1\right]$. This exemple shows a solution with a wider Kolmogorov width, requiring more modes. \cref{fig:conv_BiStiff_2D} shows the convergence of the training process and highlights the stagnation of the loss decay before each mode addition. Moreover, a global stagnation is visible at the end of the training process, where adding modes has a smaller and smaller impact. \cref{fig:2D_ROM_3_config_space_var} shows the deformed solution for three parameter configurations.

	\printbibliography[heading=bibintoc]
	
\end{document}

%% file: FENNI_PGD.tex
\tikzstyle{input} = [circle, draw=BleuLMS, fill=BleuLMS!25, minimum size=8mm, thick]

\tikzstyle{inputSpace} = [circle, circular drop shadow, draw=GreenLMS, fill=BleuLMS!25, minimum size=8mm, thick]
\tikzstyle{inputPara} = [circle, circular drop shadow, draw=LGreenLMS, fill=LGreenLMS!25, minimum size=8mm, thick]

\tikzstyle{ShapeF} = [circle, draw=GreenLMS, fill=GreenLMS!25, minimum size=2mm, thick]

\tikzstyle{Sol} = [circle, circular drop shadow, draw=LGreenLMS, fill=LGreenLMS!25, minimum size=2mm, thick]
\tikzstyle{Out} = [circle, circular drop shadow, draw=BleuLMS, fill=GreenLMS!25, minimum size=2mm, thick]
\tikzstyle{Mul} = [circle, circular drop shadow, draw=black, fill=black!25, minimum size=2mm, thick]
\tikzstyle{lineSpace} = [draw,BleuLMS, -,thick]
\tikzstyle{linePara} = [draw,LGreenLMS, -,thick]
\tikzstyle{line} = [draw,GreenLMS, -,thick]
\tikzstyle{lineDot} = [draw,GreenLMS, doted,thick]
\usetikzlibrary {math}
\usetikzlibrary {decorations}
\usetikzlibrary{shadows}

		\begin{tikzpicture}[>=Stealth, node distance=1.5cm]
		
		\def\NN{4} 
		\def\NU{3} 
		\def\NM{2} 

		\foreach \i in {1}
		\node [inputSpace] (I\i) at (0,-1.5-2.5) {$x$};
		
		\foreach \i in {2}
		\node [inputPara] (I\i) at (0,-1-1.5-6.5) {$\mu^1$};
		
		\foreach \i in {3}
		\node [inputPara] (I\i) at (0,-6-1-1.5-6.5) {$\mu^{\beta}$};
		
		\foreach \i in {2}
		\foreach \h in {3}
		\draw[loosely dotted, very thick,LGreenLMS] (I\i) -- (I\h);
		
		\foreach \h [count=\hi] in {1,...,\NN}
		\node [inputSpace] (Ho\hi) at (2, -1.1*\h) {$N^{\Omega}_\h$};
		\node [inputSpace] (Last) at (2, -1.1*\NN-1.5) {$N^{\Omega}_{n_p}$};
		\draw[dotted, very thick,GreenLMS ] (Ho\NN) -- (Last);
		
		\foreach \h [count=\hi] in {1,...,\NU}
		\node [inputPara] (Hm\hi) at (2, -1.5-4.5-1.1*\h) {$N^{\mu}_\h$};
		\node [inputPara] (Last2) at (2, -1.5-4.5-1.1*\NU-1.5) {$N^{\mu}_q$};
		\draw[ dotted, very thick, LGreenLMS] (Hm\NU) -- (Last2);
		
				\foreach \h [count=\hi] in {1,...,\NU}
				\node [inputPara] (Hmbeta\hi) at (2, -6-1.5-4.5-1.1*\h) {$N^{\mu}_\h$};
				\node [inputPara] (Last2beta) at (2, -6-1.5-4.5-1.1*\NU-1.5) {$N^{\mu}_q$};
				\draw[ dotted, very thick, LGreenLMS] (Hmbeta\NU) -- (Last2beta);
		
		\foreach \o [count=\oi] in {1,...,\NM}
		\node [inputSpace] (O\oi) at (4, -1-1.3*\oi) {$\overline{u}_\o(x)$};
		\node [inputSpace] (OLast) at (4, -1-1.3*\NM-1.7) {$\overline{u}_m(x)$};
		\draw[dotted, very thick,GreenLMS ] (O\NM) -- (OLast);
		
		\foreach \o [count=\oi] in {1,...,\NM}
		\node [inputPara] (Om\oi) at (4, -1.5-5-1.3*\oi) {$\lambda_\o^1(\mu)$};
		\node [inputPara] (OLastm) at (4, -1.5-5-1.3*\NM-1.7) {$\lambda_m(\mu)$};
		\draw[dotted, very thick,LGreenLMS ] (Om\NM) -- (OLastm);
					\foreach \o [count=\oi] in {1,...,\NM}
					\node [inputPara] (Ombeta\oi) at (4, -6-1.5-5-1.3*\oi) {$\lambda_\o^{\beta}(\mu)$};
					\node [inputPara] (OLastmbeta) at (4, -6-1.5-5-1.3*\NM-1.7) {$\lambda_m^{\beta}(\mu)$};
					\draw[dotted, very thick,LGreenLMS ] (Ombeta\NM) -- (OLastmbeta);
		
		
		\foreach \o [count=\oi] in {1,...,\NM}
		\node [Mul] (Mul\oi) at (7, -2.5-1.5-3-\oi) {$\mathcal{M}$};
		\node [Mul] (MulLast) at (7, -2.5-1.5-3-\NM-1.5) {$\mathcal{M}$};
		\draw[dotted, very thick,black ] (Mul\NM) -- (MulLast);
		
		\node [Out] (Iout) at (9,-2.5-1.5-4.5) {$u\left(x,\mu\right)$};
		
		\foreach \i in {1}
		\foreach \h in {1,...,\NN}
		\draw [lineSpace] (I\i) -- (Ho\h);
		
			\draw [lineSpace] (I1) -- (Last);
			\draw [linePara] (I2) -- (Last2);
			\draw [linePara] (I3) -- (Last2beta);
		
		\foreach \i in {2}
		\foreach \h in {1,...,\NU}
		\draw [linePara] (I\i) -- (Hm\h);

			\foreach \i in {3}
			\foreach \h in {1,...,\NU}
			\draw [linePara] (I\i) -- (Hmbeta\h);
		
		\foreach \h in {1,...,\NN}
		\foreach \o in {1,...,\NM}
		\draw [lineSpace] (Ho\h) -- (O\o);
		\draw [lineSpace] (Last) -- (O\NM);
		
		\foreach \h in {1,...,\NN}
		\draw [lineSpace] (Ho\h) -- (OLast);
		\draw [lineSpace] (Last) -- (OLast);
		
		\foreach \h in {1,...,\NM}
		\draw [lineSpace] (Last) -- (O\h);
		
		\foreach \h in {1,...,\NU}
		\foreach \o in {1,...,\NM}
		\draw [linePara] (Hm\h) -- (Om\o);
		
				\foreach \h in {1,...,\NU}
				\foreach \o in {1,...,\NM}
				\draw [linePara] (Hmbeta\h) -- (Ombeta\o);
		
		\foreach \h in {1,...,\NU}
		\draw [linePara] (Hm\h) -- (OLastm);

		\draw [linePara] (Last2) -- (OLastm);
		
		\foreach \h in {1,...,\NM}
		\draw [linePara] (Last2) -- (Om\h);
		
			\foreach \h in {1,...,\NU}
		\draw [linePara] (Hmbeta\h) -- (OLastmbeta);
				\foreach \h in {1,...,\NM}
				\draw [linePara] (Last2beta) -- (Ombeta\h);
		
		\draw [linePara] (Last2beta) -- (OLastmbeta);

		\foreach \m in {1,...,\NM}
		\draw [lineSpace] (O\m) -- (Mul\m);
		\foreach \m in {1,...,\NM}
		\draw [linePara] (Om\m) -- (Mul\m);
		\foreach \m in {1,...,\NM}
		\draw [linePara] (Ombeta\m) -- (Mul\m);
		
		\draw [lineSpace] (OLast) -- (MulLast);
		\draw [linePara] (OLastm) -- (MulLast);
		\draw [linePara] (OLastmbeta) -- (MulLast);

		\foreach \m in {1,...,\NM}
		\draw [line] (Mul\m) -- (Iout);
		
		\draw [line] (MulLast) -- (Iout);
		
	\end{tikzpicture}

%% file: Algo_greedyPGD.tex
\RestyleAlgo{ruled}

\SetKwComment{Comment}{/* }{ */}
\SetEndCharOfAlgoLine{}
\begin{algorithm}[hbtp!]
	\caption{Greedy algorithm}
	\textbf{Inputs: } \code{model}, 
	\code{optimizer},  
	\code{\arbitrary{max\_stgn}}, 
	\code{\arbitrary{$\eta_c$}}, 
	\code{\arbitrary{new\_mode\_threshold}}
	\textcolor{GreenLMS}{\Comment*[r]{Arbitrary values are highlighted in \arbitrary{red}}}

	\code{stgn = 0} 	\textcolor{GreenLMS}{\Comment*[r]{Number of consecutive stagnating epochs}}
	
	\code{$\hat{\mathcal{L}}$ = 0 } 	\textcolor{GreenLMS}{\Comment*[r]{Arbitrary initial value}}
	
	\code{usefulness = 0} 	\textcolor{GreenLMS}{\Comment*[r]{Cumulative improving epochs since new mode}}
	
	\While(\textcolor{GreenLMS}{\Comment*[r]{\hspace{-5pt}Training loop \hspace{-6pt}}}){\emph{\code{epoch}} $<$ \emph{\code{max\_epoch}}}{
		
    \If{\code{\emph{stgn}} $>$ \code{\emph{\arbitrary{max\_stgn}}} \& \textbf{\emph{not}} \code{\emph{flag\_useful}} }{
	Break\textcolor{GreenLMS}{\Comment*[r]{Break if previous mode did not stop stagnation}}
}
	$\mathcal{L} =\overline{E_p}\left(\vect{u}\left(\vect{x},\para\right),\para\right)$  \textcolor{GreenLMS}{\Comment*[r]{ Loss evaluation}}
		
    \eIf{\code{\emph{epoch>1}}}{
    \code{$\hat{\mathcal{L}}$} = $\frac{2\left(\mathcal{L}\code{\_old} - \mathcal{L}\right)}{\left|\mathcal{L}\code{\_old} + \mathcal{L}\right|}$ 	\textcolor{GreenLMS}{\Comment*[r]{Update loss decrease}}
		
		\eIf{$\left| \text{\code{\emph{$\hat{\mathcal{L}}$}}}\right| < \text{\code{\emph{\arbitrary{$\eta_c$}}}} $ }
		{\code{stgn} +=1  	\textcolor{GreenLMS}{\Comment*[r]{Increment stagnation}}
		}{
		\code{stgn}$=0$  	\textcolor{GreenLMS}{\Comment*[r]{Reset stagnation}}
		
		\If{\emph{$ \mathcal{L}$ } $> 0$}{
			\code{usefulness} +=1 \textcolor{GreenLMS}{\Comment*[r]{ Check that loss decrease is more than a single spike}}
			
			\If{\emph{\code{usefulness>\code{\arbitrary{min\_useful}}} }}{
			\code{flag\_useful = True} \textcolor{GreenLMS}{\Comment*[r]{ Previous mode flagged useful}}}  
		}
	}
    $\mathcal{L}$\code{\_old} = $\mathcal{L}$
    }{$\mathcal{L}\code{\_old} = \mathcal{L}$}
		
		\code{$\mathcal{L}$.backward()}  \textcolor{GreenLMS}{\Comment*[r]{ Gradient of the loss}}
		\code{optimiser.step()}  \textcolor{GreenLMS}{\Comment*[r]{ Update the models}}
		
			\If{\code{\emph{stgn}} $>$ \arbitrary{\code{\emph{new\_mode\_threshold}}} \& \code{\emph{flag\_useful}} }{
			\code{model.AddMode()} 	\textcolor{GreenLMS}{\Comment*[r]{Add new mode}}
			
			\code{model.AddMode2Optimizer(optimizer)} 	\textcolor{GreenLMS}{\Comment*[r]{Add the new parameters to the optimizer}}
				
			\code{flag\_useful = False} 
			
		}
	}
\label{alg:GreedyPGD}
\end{algorithm}

%% file: Body_force_1D.tex
\begin{tikzpicture}

\definecolor{darkgray176}{RGB}{176,176,176}
\definecolor{steelblue31119180}{RGB}{31,119,180}

\begin{axis}[
tick align=outside,
tick pos=left,
x grid style={darkgray176},
xlabel={$x~ \left(\mathrm{mm}\right)$},
xmin=-0.45699997805059, xmax=10.4769995193928,
xtick style={color=black},
xtick={-2,0,2,4,6,8,10,12},
xticklabels={
  $\mathdefault{\ensuremath{-}2}$,
  $\mathdefault{0}$,
  $\mathdefault{2}$,
  $\mathdefault{4}$,
  $\mathdefault{6}$,
  $\mathdefault{8}$,
  $\mathdefault{10}$,
  $\mathdefault{12}$
},
y grid style={darkgray176},
ylabel={$\vect{f}~\left(\mathrm{N}.\mathrm{mm^{-3}}\right)$},
ymin=-6.51213662624359, ymax=13.4748713254929,
ytick style={color=black},
ytick={-7.5,-5,-2.5,0,2.5,5,7.5,10,12.5,15},
yticklabels={
  $\mathdefault{\ensuremath{-}7.5}$,
  $\mathdefault{\ensuremath{-}5.0}$,
  $\mathdefault{\ensuremath{-}2.5}$,
  $\mathdefault{0.0}$,
  $\mathdefault{2.5}$,
  $\mathdefault{5.0}$,
  $\mathdefault{7.5}$,
  $\mathdefault{10.0}$,
  $\mathdefault{12.5}$,
  $\mathdefault{15.0}$
}
]
\addplot [very thick, steelblue31119180]
table {%
0.0399999991059303 -1.28823387512966e-06
0.0599999986588955 -1.72353202287923e-06
0.0799999982118607 -2.29979832511162e-06
0.100000001490116 -3.06058745991322e-06
0.119999997317791 -4.06220351578668e-06
0.140000000596046 -5.37727783012087e-06
0.159999996423721 -7.09907817508793e-06
0.180000007152557 -9.3471917352872e-06
0.200000002980232 -1.22743140309467e-05
0.219999998807907 -1.60750114446273e-05
0.239999994635582 -2.0996176317567e-05
0.259999990463257 -2.7350286472938e-05
0.280000001192093 -3.55317642970476e-05
0.300000011920929 -4.60364972241223e-05
0.319999992847443 -5.94862576690502e-05
0.340000003576279 -7.66582234064117e-05
0.360000014305115 -9.85208607744426e-05
0.379999995231628 -0.000126276005175896
0.400000005960464 -0.000161412099259906
0.419999986886978 -0.000205764721613377
0.439999997615814 -0.000261592620518059
0.46000000834465 -0.000331662915414199
0.479999989271164 -0.000419355259509757
0.5 -0.00052878720453009
0.519999980926514 -0.000664951629005373
0.540000021457672 -0.000833887024782598
0.560000002384186 -0.00104287185240537
0.579999983310699 -0.00130064249970019
0.600000023841858 -0.00161765760276467
0.620000004768372 -0.00200637988746166
0.639999985694885 -0.00248162215575576
0.660000026226044 -0.00306091853417456
0.680000007152557 -0.0037649346049875
0.699999988079071 -0.00461796252056956
0.720000028610229 -0.005648422986269
0.740000009536743 -0.00688945734873414
0.759999990463257 -0.00837954320013523
0.779999971389771 -0.0101631730794907
0.800000011920929 -0.0122916093096137
0.819999992847443 -0.0148236230015755
0.839999973773956 -0.0178263708949089
0.860000014305115 -0.0213761664927006
0.879999995231628 -0.0255593899637461
0.899999976158142 -0.0304733514785767
0.920000016689301 -0.0362272337079048
0.939999997615814 -0.0429426841437817
0.959999978542328 -0.0507548153400421
0.980000019073486 -0.0598128251731396
1 -0.0702804923057556
1.01999998092651 -0.0823363810777664
1.03999996185303 -0.0961744636297226
1.05999994277954 -0.112003549933434
1.08000004291534 -0.130047038197517
1.10000002384186 -0.150542214512825
1.12000000476837 -0.173739165067673
1.13999998569489 -0.199898704886436
1.1599999666214 -0.229290887713432
1.17999994754791 -0.262191295623779
1.20000004768372 -0.298878699541092
1.22000002861023 -0.339629679918289
1.24000000953674 -0.384715139865875
1.25999999046326 -0.434394270181656
1.27999997138977 -0.4889075756073
1.29999995231628 -0.548471689224243
1.32000005245209 -0.613269746303558
1.3400000333786 -0.683443427085876
1.36000001430511 -0.759085357189178
1.37999999523163 -0.840228199958801
1.39999997615814 -0.926836252212524
1.41999995708466 -1.01879346370697
1.44000005722046 -1.11589682102203
1.46000003814697 -1.21784138679504
1.48000001907349 -1.3242175579071
1.5 -1.43449592590332
1.51999998092651 -1.54802477359772
1.53999996185303 -1.66402161121368
1.55999994277954 -1.78156661987305
1.58000004291534 -1.89960467815399
1.60000002384186 -2.01693773269653
1.62000000476837 -2.13223361968994
1.63999998569489 -2.24402642250061
1.6599999666214 -2.35072684288025
1.67999994754791 -2.4506299495697
1.70000004768372 -2.54193234443665
1.72000002861023 -2.62274742126465
1.74000000953674 -2.69112801551819
1.75999999046326 -2.7450909614563
1.77999997138977 -2.7826452255249
1.79999995231628 -2.80181813240051
1.82000005245209 -2.80069446563721
1.8400000333786 -2.77744936943054
1.86000001430511 -2.73038291931152
1.87999999523163 -2.65796184539795
1.89999997615814 -2.55885314941406
1.91999995708466 -2.43196702003479
1.94000005722046 -2.27648711204529
1.96000003814697 -2.09191107749939
1.98000001907349 -1.87807512283325
2 -1.63518500328064
2.01999998092651 -1.36383616924286
2.03999996185303 -1.06503117084503
2.05999994277954 -0.740190863609314
2.07999992370605 -0.391155868768692
2.09999990463257 -0.0201825220137835
2.11999988555908 0.370065659284592
2.14000010490417 0.776547491550446
2.16000008583069 1.19584810733795
2.1800000667572 1.62424576282501
2.20000004768372 2.05773448944092
2.22000002861023 2.49207925796509
2.24000000953674 2.92287158966064
2.25999999046326 3.34559035301208
2.27999997138977 3.75566625595093
2.29999995231628 4.14855337142944
2.3199999332428 4.51979494094849
2.33999991416931 4.86509799957275
2.35999989509583 5.18039751052856
2.38000011444092 5.4619312286377
2.40000009536743 5.70628833770752
2.42000007629395 5.91048431396484
2.44000005722046 6.07200145721436
2.46000003814697 6.18883323669434
2.48000001907349 6.25952339172363
2.5 6.28318548202515
2.51999998092651 6.25952339172363
2.53999996185303 6.18883323669434
2.55999994277954 6.07200145721436
2.57999992370605 5.91048431396484
2.59999990463257 5.70628833770752
2.61999988555908 5.4619312286377
2.64000010490417 5.18039751052856
2.66000008583069 4.86509799957275
2.6800000667572 4.51979494094849
2.70000004768372 4.14855337142944
2.72000002861023 3.75566625595093
2.74000000953674 3.34559035301208
2.75999999046326 2.92287158966064
2.77999997138977 2.49207925796509
2.79999995231628 2.05773448944092
2.8199999332428 1.62424576282501
2.83999991416931 1.19584810733795
2.85999989509583 0.776547491550446
2.88000011444092 0.370065659284592
2.90000009536743 -0.0201825220137835
2.92000007629395 -0.391155868768692
2.94000005722046 -0.740190863609314
2.96000003814697 -1.06503117084503
2.98000001907349 -1.36383616924286
3 -1.63518500328064
3.01999998092651 -1.87807512283325
3.03999996185303 -2.09191107749939
3.05999994277954 -2.27648711204529
3.07999992370605 -2.43196606636047
3.09999990463257 -2.5588526725769
3.11999988555908 -2.65796136856079
3.14000010490417 -2.73038339614868
3.16000008583069 -2.7774498462677
3.1800000667572 -2.80069518089294
3.20000004768372 -2.80181813240051
3.22000002861023 -2.7826452255249
3.24000000953674 -2.7450909614563
3.25999999046326 -2.69112801551819
3.27999997138977 -2.62274742126465
3.29999995231628 -2.54193234443665
3.3199999332428 -2.45063090324402
3.33999991416931 -2.35072755813599
3.35999989509583 -2.24402761459351
3.38000011444092 -2.1322329044342
3.40000009536743 -2.0169370174408
3.42000007629395 -1.89960384368896
3.44000005722046 -1.78156661987305
3.46000003814697 -1.66402161121368
3.48000001907349 -1.54802477359772
3.5 -1.43449592590332
3.51999998092651 -1.3242175579071
3.53999996185303 -1.21784138679504
3.55999994277954 -1.11589682102203
3.57999992370605 -1.01879405975342
3.59999990463257 -0.926836669445038
3.61999988555908 -0.840228617191315
3.64000010490417 -0.759084820747375
3.66000008583069 -0.683443009853363
3.6800000667572 -0.6132692694664
3.70000004768372 -0.548471689224243
3.72000002861023 -0.4889075756073
3.74000000953674 -0.434394270181656
3.75999999046326 -0.384715139865875
3.77999997138977 -0.339629679918289
3.79999995231628 -0.298878699541092
3.8199999332428 -0.262191474437714
3.83999991416931 -0.229291066527367
3.85999989509583 -0.199898943305016
3.88000011444092 -0.173739045858383
3.90000009536743 -0.150542110204697
3.92000007629395 -0.130046993494034
3.94000005722046 -0.112003549933434
3.96000003814697 -0.0961744636297226
3.98000001907349 -0.0823363810777664
4 -0.0702804923057556
4.01999998092651 -0.0598128251731396
4.03999996185303 -0.0507548153400421
4.05999994277954 -0.0429426841437817
4.07999992370605 -0.0362272337079048
4.09999990463257 -0.0304733756929636
4.11999988555908 -0.0255594104528427
4.1399998664856 -0.0213762056082487
4.15999984741211 -0.0178264155983925
4.17999982833862 -0.0148236602544785
4.19999980926514 -0.0122916409745812
4.21999979019165 -0.0101632000878453
4.23999977111816 -0.00837956462055445
4.26000022888184 -0.00688944524154067
4.28000020980835 -0.00564840855076909
4.30000019073486 -0.00461795087903738
4.32000017166138 -0.00376492482610047
4.34000015258789 -0.00306091038510203
4.3600001335144 -0.00248162006027997
4.38000011444092 -0.00200637616217136
4.40000009536743 -0.00161765480879694
4.42000007629395 -0.00130064261611551
4.44000005722046 -0.00104287196882069
4.46000003814697 -0.000833887199405581
4.48000001907349 -0.000664951861836016
4.5 -0.000528787553776056
4.51999998092651 -0.000419355783378705
4.53999996185303 -0.000331663672113791
4.55999994277954 -0.000261593697359785
4.57999992370605 -0.000205766264116392
4.59999990463257 -0.00016141428204719
4.61999988555908 -0.00012627910473384
4.6399998664856 -9.85252409009263e-05
4.65999984741211 -7.6664611697197e-05
4.67999982833862 -5.94951052335091e-05
4.69999980926514 -4.60488081444055e-05
4.71999979019165 -3.55489210051019e-05
4.73999977111816 -2.73741698038066e-05
4.76000022888184 -2.10292055271566e-05
4.78000020980835 -1.61209663929185e-05
4.80000019073486 -1.23380068544066e-05
4.82000017166138 -9.43514805840096e-06
4.84000015258789 -7.22024924471043e-06
4.8600001335144 -5.54375583305955e-06
4.88000011444092 -4.29033207183238e-06
4.90000009536743 -3.37233677782933e-06
4.92000007629395 -2.72469583251223e-06
4.94000005722046 -2.30111686505552e-06
4.96000003814697 -2.071300514217e-06
4.98000001907349 -2.01916418518522e-06
5 -2.14193437386712e-06
5.01999998092651 -2.45006276600179e-06
5.03999996185303 -2.96800089927274e-06
5.05999994277954 -3.73585635315976e-06
5.07999992370605 -4.81204551761039e-06
5.09999990463257 -6.2770495787845e-06
5.11999988555908 -8.23847130959621e-06
5.1399998664856 -1.0837765330507e-05
5.15999984741211 -1.42587032314623e-05
5.17999982833862 -1.87383102456806e-05
5.19999980926514 -2.45804440055508e-05
5.21999979019165 -3.21729130519088e-05
5.23999977111816 -4.20086944359355e-05
5.26000022888184 -5.47126219316851e-05
5.28000020980835 -7.10722451913171e-05
5.30000019073486 -9.20793318073265e-05
5.32000017166138 -0.000118977171950974
5.34000015258789 -0.000153319953824393
5.3600001335144 -0.000197043918888085
5.38000011444092 -0.000252553552854806
5.40000009536743 -0.000322825304465368
5.42000007629395 -0.000411530199926347
5.44000005722046 -0.000523185764905065
5.46000003814697 -0.000663326180074364
5.48000001907349 -0.000838710810057819
5.5 -0.00105757464189082
5.51999998092651 -0.00132990337442607
5.53999996185303 -0.00166777416598052
5.55999994277954 -0.00208574370481074
5.57999992370605 -0.00260128499940038
5.59999990463257 -0.00323530961759388
5.61999988555908 -0.00401275232434273
5.6399998664856 -0.00496324012055993
5.65999984741211 -0.00612182077020407
5.67999982833862 -0.00752984965220094
5.69999980926514 -0.00923590175807476
5.71999979019165 -0.0112968171015382
5.73999977111816 -0.0137788904830813
5.76000022888184 -0.0167591292411089
5.78000020980835 -0.0203264001756907
5.80000019073486 -0.0245832819491625
5.82000017166138 -0.0296473205089569
5.84000015258789 -0.035652831196785
5.8600001335144 -0.0427524112164974
5.88000011444092 -0.0511188209056854
5.90000009536743 -0.0609467513859272
5.92000007629395 -0.0724544674158096
5.94000005722046 -0.0858853682875633
5.96000003814697 -0.101509630680084
5.98000001907349 -0.119625650346279
6 -0.140560984611511
6.01999998092651 -0.164672762155533
6.03999996185303 -0.192348927259445
6.05999994277954 -0.224007099866867
6.07999992370605 -0.260093986988068
6.09999990463257 -0.301084220409393
6.11999988555908 -0.347478091716766
6.1399998664856 -0.399797320365906
6.15999984741211 -0.458581447601318
6.17999982833862 -0.524382412433624
6.19999980926514 -0.597756505012512
6.21999979019165 -0.679258346557617
6.23999977111816 -0.7694291472435
6.26000022888184 -0.868789792060852
6.28000020980835 -0.977816581726074
6.30000019073486 -1.09694457054138
6.32000017166138 -1.22654032707214
6.34000015258789 -1.36688792705536
6.3600001335144 -1.51817166805267
6.38000011444092 -1.68045723438263
6.40000009536743 -1.85367333889008
6.42000007629395 -2.03758811950684
6.44000005722046 -2.23179364204407
6.46000003814697 -2.43568277359009
6.48000001907349 -2.64843511581421
6.5 -2.86899185180664
6.51999998092651 -3.09604954719543
6.53999996185303 -3.32804322242737
6.55999994277954 -3.56313323974609
6.57999992370605 -3.79920768737793
6.59999990463257 -4.03387403488159
6.61999988555908 -4.26446580886841
6.6399998664856 -4.4880518913269
6.65999984741211 -4.70145273208618
6.67999982833862 -4.90125894546509
6.69999980926514 -5.0838623046875
6.71999979019165 -5.24549293518066
6.73999977111816 -5.38225412368774
6.76000022888184 -5.49018335342407
6.78000020980835 -5.56528997421265
6.80000019073486 -5.60363626480103
6.82000017166138 -5.60138893127441
6.84000015258789 -5.55489826202393
6.8600001335144 -5.46076536178589
6.88000011444092 -5.31592273712158
6.90000009536743 -5.11770534515381
6.92000007629395 -4.86393213272095
6.94000005722046 -4.55297422409058
6.96000003814697 -4.18382215499878
6.98000001907349 -3.7561502456665
7 -3.27037000656128
7.01999998092651 -2.72767233848572
7.03999996185303 -2.13006234169006
7.05999994277954 -1.48038172721863
7.07999992370605 -0.782311737537384
7.09999990463257 -0.0403650440275669
7.11999988555908 0.740131318569183
7.1399998664856 1.55308520793915
7.15999984741211 2.39168643951416
7.17999982833862 3.2484815120697
7.19999980926514 4.11545896530151
7.21999979019165 4.9841480255127
7.23999977111816 5.84573316574097
7.26000022888184 6.69118976593018
7.28000020980835 7.51134252548218
7.30000019073486 8.29711532592773
7.32000017166138 9.03959941864014
7.34000015258789 9.73020362854004
7.3600001335144 10.360800743103
7.38000011444092 10.9238624572754
7.40000009536743 11.412576675415
7.42000007629395 11.8209686279297
7.44000005722046 12.1440029144287
7.46000003814697 12.3776664733887
7.48000001907349 12.5190467834473
7.5 12.5663709640503
7.51999998092651 12.5190467834473
7.53999996185303 12.3776664733887
7.55999994277954 12.1440029144287
7.57999992370605 11.8209686279297
7.59999990463257 11.412576675415
7.61999988555908 10.9238624572754
7.6399998664856 10.360800743103
7.65999984741211 9.73020362854004
7.67999982833862 9.03959941864014
7.69999980926514 8.29711532592773
7.71999979019165 7.51134252548218
7.73999977111816 6.69118976593018
7.76000022888184 5.84573316574097
7.78000020980835 4.9841480255127
7.80000019073486 4.11545896530151
7.82000017166138 3.2484815120697
7.84000015258789 2.39168643951416
7.8600001335144 1.55308520793915
7.88000011444092 0.740131318569183
7.90000009536743 -0.0403650440275669
7.92000007629395 -0.782311737537384
7.94000005722046 -1.48038172721863
7.96000003814697 -2.13006234169006
7.98000001907349 -2.72767233848572
8 -3.27037000656128
8.02000045776367 -3.75616121292114
8.03999996185303 -4.18382215499878
8.0600004196167 -4.55298233032227
8.07999992370605 -4.86393213272095
8.10000038146973 -5.11771011352539
8.11999988555908 -5.31592273712158
8.14000034332275 -5.46076774597168
8.15999984741211 -5.55489826202393
8.18000030517578 -5.60138940811157
8.19999980926514 -5.60363626480103
8.22000026702881 -5.56528902053833
8.23999977111816 -5.49018335342407
8.26000022888184 -5.38225412368774
8.27999973297119 -5.24549627304077
8.30000019073486 -5.0838623046875
8.31999969482422 -4.90126323699951
8.34000015258789 -4.70145273208618
8.35999965667725 -4.48805713653564
8.38000011444092 -4.26446580886841
8.39999961853027 -4.03387928009033
8.42000007629395 -3.79920768737793
8.4399995803833 -3.56313920021057
8.46000003814697 -3.32804322242737
8.47999954223633 -3.09605550765991
8.5 -2.86899185180664
8.52000045776367 -2.64842963218689
8.53999996185303 -2.43568277359009
8.5600004196167 -2.23178911209106
8.57999992370605 -2.03758811950684
8.60000038146973 -1.85366904735565
8.61999988555908 -1.68045723438263
8.64000034332275 -1.51816749572754
8.65999984741211 -1.36688792705536
8.68000030517578 -1.22653675079346
8.69999980926514 -1.09694457054138
8.72000026702881 -0.977814137935638
8.73999977111816 -0.868789792060852
8.76000022888184 -0.7694291472435
8.77999973297119 -0.679260432720184
8.80000019073486 -0.597756505012512
8.81999969482422 -0.524383783340454
8.84000015258789 -0.458581447601318
8.85999965667725 -0.399798542261124
8.88000011444092 -0.347478091716766
8.89999961853027 -0.301085293292999
8.92000007629395 -0.260093986988068
8.9399995803833 -0.224007695913315
8.96000003814697 -0.192348927259445
8.97999954223633 -0.164673432707787
9 -0.140560984611511
9.02000045776367 -0.119625210762024
9.03999996185303 -0.101509630680084
9.0600004196167 -0.0858849808573723
9.07999992370605 -0.0724544674158096
9.10000038146973 -0.0609464980661869
9.11999988555908 -0.0511188209056854
9.14000034332275 -0.0427521951496601
9.15999984741211 -0.035652831196785
9.18000030517578 -0.0296471994370222
9.19999980926514 -0.0245832819491625
9.22000026702881 -0.0203262958675623
9.23999977111816 -0.0167591292411089
9.26000022888184 -0.0137788904830813
9.27999973297119 -0.0112968757748604
9.30000019073486 -0.00923590175807476
9.31999969482422 -0.00752988876774907
9.34000015258789 -0.00612182077020407
9.35999965667725 -0.00496326154097915
9.38000011444092 -0.00401275232434273
9.39999961853027 -0.0032353294081986
9.42000007629395 -0.00260128499940038
9.4399995803833 -0.00208575441502035
9.46000003814697 -0.0016677740495652
9.47999954223633 -0.00132991024293005
9.5 -0.00105757440906018
9.52000045776367 -0.000838706095237285
9.53999996185303 -0.000663325830828398
9.5600004196167 -0.000523181981407106
9.57999992370605 -0.000411529443226755
9.60000038146973 -0.000322821899317205
9.61999988555908 -0.000252552010351792
9.64000034332275 -0.000197040688362904
9.65999984741211 -0.000153316868818365
9.68000030517578 -0.000118972086056601
9.69999980926514 -9.20732345548458e-05
9.72000026702881 -7.10633394191973e-05
9.73999977111816 -5.47007184650283e-05
9.76000022888184 -4.19921234424692e-05
9.77999973297119 -3.21501356665976e-05
9.80000019073486 -2.45485880441265e-05
9.81999969482422 -1.86944507731823e-05
9.84000015258789 -1.41981045089778e-05
9.85999965667725 -1.07545947685139e-05
9.88000011444092 -8.12440703157336e-06
9.89999961853027 -6.12120811638306e-06
9.92000007629395 -4.59959665022325e-06
9.9399995803833 -3.44708882948908e-06
9.96000003814697 -2.57646775025933e-06
9.97999954223633 -1.92065522242046e-06
};
\end{axis}

\end{tikzpicture}

%% file: L2_1D_Mono.tex
\begin{tikzpicture}
	
	\begin{axis}[
		xlabel={Epochs},
		ylabel=\textcolor{BleuLMS}{Error - $\mathcal{\eta_{\text{A}}}$},
		ymode=log,
		axis y line*=left,
		xmajorgrids=true,
		ymajorgrids=true,
		ytick style={color=BleuLMS},
		yticklabel style={anchor=east,color=BleuLMS},
		ytick={1e-07,1e-06,1e-05,1e-04,1e-03,1e-02,1e-01,1},
		yticklabels={
						$\mathdefault{10^{-7}}$,
						$\mathdefault{10^{-6}}$,
			$\mathdefault{10^{-5}}$,
			$\mathdefault{10^{-4}}$,
			$\mathdefault{10^{-3}}$,
			$\mathdefault{10^{-2}}$,
			$\mathdefault{10^{-1}}$,
			$\mathdefault{10^{0}}$
		},
		]
		\addplot[
		color=BleuLMS,
		very thick,
		]
		table [col sep=comma, x=epochs, y=Error] {Fig5.csv};
		
		\end{axis}
	
	\begin{axis}[
		xlabel={Epochs},
		ylabel=\textcolor{RougeLMS}{Modes},
		axis y line*=right,
		axis x line=none,
		ytick style={color=RougeLMS},
		yticklabel style={color=RougeLMS},
		ytick={1,2},
		ymajorgrids=true,
		]
		\addplot[
		color=RougeLMS,
		very thick,
		]
		table [col sep=comma, x=epochs, y=Modes] {Fig5.csv};
		
	\end{axis}
	
\end{tikzpicture}

%% file: Loss_1D_Mono.tex
\begin{tikzpicture}
	
    \begin{axis}[
	xlabel={Epochs},
	ylabel=\textcolor{BleuLMS}{$\mathcal{L}$oss},
	y dir=reverse,
	ymode=log,
	axis y line*=left,
	xmajorgrids=true,
	ymajorgrids=true,
	ytick style={color=BleuLMS},
	yticklabel style={anchor=east,color=BleuLMS},
	ytick={1e-05,1e-04,1e-03,1e-02,1e-01,1},
	yticklabels={
			$-\mathdefault{10^{-5}}$,
			$-\mathdefault{10^{-4}}$,
			$-\mathdefault{10^{-3}}$,
			$-\mathdefault{10^{-2}}$,
			$-\mathdefault{10^{-1}}$,
			$-\mathdefault{10^{0}}$
		},
	]
	\addplot[
	color=BleuLMS,
	very thick,
	]
	table [col sep=comma, x=epochs, y=Loss] {Fig5.csv};

x\end{axis}

\begin{axis}[
	xlabel={Epochs},
	ylabel=\textcolor{RougeLMS}{Modes},
	axis y line*=right,
	axis x line=none,
	ytick style={color=RougeLMS},
	yticklabel style={color=RougeLMS},
	ytick={1,2},
	ymajorgrids=true,
	]
	\addplot[
	color=RougeLMS,
	very thick,
	]
	table [col sep=comma, x=epochs, y=Modes] {Fig5.csv};

\end{axis}
	
\end{tikzpicture}

%% file: LossDecay_1D_Mono.tex
\begin{tikzpicture}
	
	\begin{axis}[
		xlabel={Epochs},
		ylabel=\textcolor{BleuLMS}{Loss decay},
		ymode=log,
		axis y line*=left,
		xmajorgrids=true,
		ymajorgrids=true,
		ytick style={color=BleuLMS},
		yticklabel style={anchor=east,color=BleuLMS},
		ytick={1e-07,1e-06,1e-05,1e-04,1e-03,1e-02,1e-01,1},
yticklabels={
	$\mathdefault{10^{-7}}$,
	$\mathdefault{10^{-6}}$,
	$\mathdefault{10^{-5}}$,
	$\mathdefault{10^{-4}}$,
	$\mathdefault{10^{-3}}$,
	$\mathdefault{10^{-2}}$,
	$\mathdefault{10^{-1}}$,
	$\mathdefault{10^{0}}$
},
		]
		\addplot[
		color=BleuLMS,
		very thick,
		]
		table [col sep=comma, x=epochs, y=Decay] {Fig5.csv};

							\addplot [very thick, black] table[col sep=comma, x=epochs, y expr=1e-5] {Fig5.csv};
		\end{axis}
	
	\begin{axis}[
		xlabel={Epochs},
		ylabel=\textcolor{RougeLMS}{Modes},
		axis y line*=right,
		axis x line=none,
		ytick style={color=RougeLMS},
		yticklabel style={color=RougeLMS},
		ytick={1,2},
		ymajorgrids=true,
		]
		\addplot[
		color=RougeLMS,
		very thick,
		]
		table [col sep=comma, x=epochs, y=Modes] {Fig5.csv};

	\end{axis}

\end{tikzpicture}

%% file: Loss_1D_Bi_seq.tex
\begin{tikzpicture}
	
	\begin{axis}[
		xlabel={Epochs},
		ylabel=\textcolor{BleuLMS}{Loss},
	y dir=reverse,
		ymode=log,
		axis y line*=left,
		xmajorgrids=true,
		ymajorgrids=true,
		ytick style={color=BleuLMS},
		yticklabel style={anchor=east,color=BleuLMS},
		ytick={1e-07,1e-06,1e-05,1e-04,1e-03,1e-02,1e-01,1},
		yticklabels={
			$-\mathdefault{10^{-7}}$,
			$-\mathdefault{10^{-6}}$,
			$-\mathdefault{10^{-5}}$,
			$-\mathdefault{10^{-4}}$,
			$-\mathdefault{10^{-3}}$,
			$-\mathdefault{10^{-2}}$,
			$-\mathdefault{10^{-1}}$,
			$-\mathdefault{10^{0}}$
		},
		restrict x to domain=0:2000,
		filter discard warning=false,
		]
		\addplot[
		color=BleuLMS,
		very thick,
		]
		table [col sep=comma, x=epochs, y=Loss] {Fig6.csv};
		
		\end{axis}
	
	\begin{axis}[
		xlabel={Epochs},
		ylabel=\textcolor{RougeLMS}{Modes},
		axis y line*=right,
		axis x line=none,
		ytick style={color=RougeLMS},
		yticklabel style={color=RougeLMS},
		ytick={1,2,3,4},
		restrict x to domain=0:2000,
		filter discard warning=false,
		]
		\addplot[
		color=RougeLMS,
		very thick,
		]
		table [col sep=comma, x=epochs, y=Modes] {Fig6.csv};
		
	\end{axis}
	
\end{tikzpicture}

%% file: Loss_1D_Bi_seq_zoom.tex
\begin{tikzpicture}
	
	\begin{axis}[
		xlabel={Epochs},
		ylabel=\textcolor{BleuLMS}{Loss},
		y dir=reverse,
		ymode=log,
		axis y line*=left,
		xmajorgrids=true,
		ymajorgrids=true,
		ytick style={color=BleuLMS},
		yticklabel style={anchor=east,color=BleuLMS},
		ytick={0.036,0.037},
		yticklabels={
			$-3.6\times\mathdefault{10^{-2}}$,
			$-3.7\times \mathdefault{10^{-2}}$
		},
		ymin=0.036, ymax=0.0375,
		filter discard warning=false,
		]
		\addplot[
		color=BleuLMS,
		very thick,
		]
		table [col sep=comma, x=epochs_truncated, y=Loss_truncated] {Fig6.csv};
		
	\end{axis}
	
	\begin{axis}[
		xlabel={Epochs},
		ylabel=\textcolor{RougeLMS}{Modes},
		axis y line*=right,
		axis x line=none,
		ytick style={color=RougeLMS},
		yticklabel style={color=RougeLMS},
		ytick={1,2,3,4},
		restrict y to domain=0:100,
		filter discard warning=false,
		]
		\addplot[
		color=RougeLMS,
		very thick,
		]
		table [col sep=comma, x=epochs_truncated, y=Modes_truncated] {Fig6.csv};
		
	\end{axis}
	
\end{tikzpicture}

%% file: LossDecay_1D_Bi_seq.tex
\begin{tikzpicture}
	
	\begin{axis}[
		xlabel={Epochs},
		ylabel=\textcolor{BleuLMS}{Loss decay},
		ymode=log,
		axis y line*=left,
		xmajorgrids=true,
		ymajorgrids=true,
		ytick style={color=BleuLMS},
		yticklabel style={anchor=east,color=BleuLMS},
		ytick={1e-07,1e-06,1e-05,1e-04,1e-03,1e-02,1e-01,1},
yticklabels={
	$\mathdefault{10^{-7}}$,
	$\mathdefault{10^{-6}}$,
	$\mathdefault{10^{-5}}$,
	$\mathdefault{10^{-4}}$,
	$\mathdefault{10^{-3}}$,
	$\mathdefault{10^{-2}}$,
	$\mathdefault{10^{-1}}$,
	$\mathdefault{10^{0}}$
},
		]
		\addplot[
		color=BleuLMS,
		very thick,
		]
		table [col sep=comma, x=epochs, y=Decay] {Fig6.csv};
		\addplot [very thick, black] table[col sep=comma, x=epochs, y expr=1e-5] {Fig6.csv};
		\end{axis}
	
	\begin{axis}[
		xlabel={Epochs},
		ylabel=\textcolor{RougeLMS}{Modes},
		axis y line*=right,
		axis x line=none,
		ytick style={color=RougeLMS},
		yticklabel style={color=RougeLMS},
		ytick={1,2,3,4},
		]
		\addplot[
		color=RougeLMS,
		very thick,
		]
		table [col sep=comma, x=epochs, y=Modes] {Fig6.csv};

	\end{axis}

\end{tikzpicture}

%% file: Loss_1D_Bi_zoom.tex
\begin{tikzpicture}
	
	\begin{axis}[
		xlabel={Epochs},
		ylabel=\textcolor{BleuLMS}{Loss},
		y dir=reverse,
		ymode=log,
		axis y line*=left,
		xmajorgrids=true,
		ymajorgrids=true,
		ytick style={color=BleuLMS},
		yticklabel style={anchor=east,color=BleuLMS},
		ytick={0.036,0.037},
		yticklabels={
			$-3.6\times\mathdefault{10^{-2}}$,
			$-3.7\times \mathdefault{10^{-2}}$
		},
		ymin=0.036, ymax=0.0375,
		filter discard warning=false,
		]
		\addplot[
		color=BleuLMS,
		very thick,
		]
		table [col sep=comma, x=epochs_truncated, y=Loss_truncated] {Fig7.csv};
		
	\end{axis}
	
	\begin{axis}[
		xlabel={Epochs},
		ylabel=\textcolor{RougeLMS}{Modes},
		axis y line*=right,
		axis x line=none,
		ytick style={color=RougeLMS},
		yticklabel style={color=RougeLMS},
		ytick={1,2,3},
		restrict y to domain=0:100,
		filter discard warning=false,
		]
		\addplot[
		color=RougeLMS,
		very thick,
		]
		table [col sep=comma, x=epochs_truncated, y=Modes_truncated] {Fig7.csv};
		
	\end{axis}
	
\end{tikzpicture}

%% file: LossDecay_1D_Bi.tex
\begin{tikzpicture}
	
	\begin{axis}[
		xlabel={Epochs},
		ylabel=\textcolor{BleuLMS}{Loss decay},
		ymode=log,
		axis y line*=left,
		xmajorgrids=true,
		ymajorgrids=true,
		ytick style={color=BleuLMS},
		yticklabel style={anchor=east,color=BleuLMS},
		ytick={1e-07,1e-06,1e-05,1e-04,1e-03,1e-02,1e-01,1},
yticklabels={
	$\mathdefault{10^{-7}}$,
	$\mathdefault{10^{-6}}$,
	$\mathdefault{10^{-5}}$,
	$\mathdefault{10^{-4}}$,
	$\mathdefault{10^{-3}}$,
	$\mathdefault{10^{-2}}$,
	$\mathdefault{10^{-1}}$,
	$\mathdefault{10^{0}}$
},
		]
		\addplot[
		color=BleuLMS,
		very thick,
		]
		table [col sep=comma, x=epochs, y=Decay] {Fig7.csv};
		\addplot [very thick, black] table[col sep=comma, x=epochs, y expr=1e-5] {Fig7.csv};
		\end{axis}
	
	\begin{axis}[
		xlabel={Epochs},
		ylabel=\textcolor{RougeLMS}{Modes},
		axis y line*=right,
		axis x line=none,
		ytick style={color=RougeLMS},
		yticklabel style={color=RougeLMS},
		ytick={1,2,3},
		]
		\addplot[
		color=RougeLMS,
		very thick,
		]
		table [col sep=comma, x=epochs, y=Modes] {Fig7.csv};

	\end{axis}

\end{tikzpicture}

%% file: Loss_2D_Bi.tex
\begin{tikzpicture}
	
	\begin{axis}[
		xlabel={Epochs},
		ylabel=\textcolor{BleuLMS}{Loss},
	y dir=reverse,
		ymode=log,
		axis y line*=left,
		xmajorgrids=true,
		ymajorgrids=true,
		ytick style={color=BleuLMS},
		yticklabel style={anchor=east,color=BleuLMS},
		ytick={1e-07,1e-06,1e-05,1e-04,1e-03,1e-02,1e-01,1},
		yticklabels={
			$-\mathdefault{10^{-7}}$,
$-\mathdefault{10^{-6}}$,
$-\mathdefault{10^{-5}}$,
$-\mathdefault{10^{-4}}$,
$-\mathdefault{10^{-3}}$,
$-\mathdefault{10^{-2}}$,
$-\mathdefault{10^{-1}}$,
$-\mathdefault{10^{0}}$
		},
		restrict x to domain=0:2000,
		filter discard warning=false,
		]
		\addplot[
		color=BleuLMS,
		very thick,
		]
		table [col sep=comma, x=epochs, y=Loss] {Fig8.csv};
		
		\end{axis}
	
	\begin{axis}[
		xlabel={Epochs},
		ylabel=\textcolor{RougeLMS}{Modes},
		axis y line*=right,
		axis x line=none,
		ytick style={color=RougeLMS},
		yticklabel style={color=RougeLMS},
		ytick={1,2,3},
		restrict x to domain=0:2000,
		filter discard warning=false,
		]
		\addplot[
		color=RougeLMS,
		very thick,
		]
		table [col sep=comma, x=epochs, y=Modes] {Fig8.csv};
		
	\end{axis}
	
\end{tikzpicture}

%% file: Loss_2D_Bi_zoom.tex
\begin{tikzpicture}
	
	\begin{axis}[
		xlabel={Epochs},
		ylabel=\textcolor{BleuLMS}{Loss},
		y dir=reverse,
		ymode=log,
		axis y line*=left,
		xmajorgrids=true,
		ymajorgrids=true,
		ytick style={color=BleuLMS},
		yticklabel style={anchor=east,color=BleuLMS},
		ytick={0.004,0.006},
		yticklabels={
  $\mathdefault{-4\mathrm{e}{-3}}$,
$\mathdefault{-6\mathrm{e}{-3}}$
		},
		ymin=0.0039, ymax=0.0061,
		filter discard warning=false,
		]
		\addplot[
		color=BleuLMS,
		very thick,
		]
		table [col sep=comma, x=epochs_truncated, y=Loss_truncated] {Fig8.csv};
		
	\end{axis}
	
	\begin{axis}[
		xlabel={Epochs},
		ylabel=\textcolor{RougeLMS}{Modes},
		axis y line*=right,
		axis x line=none,
		ytick style={color=RougeLMS},
		yticklabel style={color=RougeLMS},
		ytick={1,2,3},
		ymajorgrids=true,
		restrict y to domain=0:100,
		filter discard warning=false,
		]
		\addplot[
		color=RougeLMS,
		very thick,
		]
		table [col sep=comma, x=epochs_truncated, y=Modes_truncated] {Fig8.csv};
		
	\end{axis}
	
\end{tikzpicture}

%% file: LossDecay_2D_Bi.tex
\begin{tikzpicture}
	
	\begin{axis}[
		xlabel={Epochs},
		ylabel=\textcolor{BleuLMS}{Loss decay},
		ymode=log,
		axis y line*=left,
		xmajorgrids=true,
		ymajorgrids=true,
		ytick style={color=BleuLMS},
		yticklabel style={anchor=east,color=BleuLMS},
		ytick={1e-07,1e-06,1e-05,1e-04,1e-03,1e-02,1e-01,1},
yticklabels={
	$\mathdefault{10^{-7}}$,
	$\mathdefault{10^{-6}}$,
	$\mathdefault{10^{-5}}$,
	$\mathdefault{10^{-4}}$,
	$\mathdefault{10^{-3}}$,
	$\mathdefault{10^{-2}}$,
	$\mathdefault{10^{-1}}$,
	$\mathdefault{10^{0}}$
},
		]
		\addplot[
		color=BleuLMS,
		very thick,
		]
		table [col sep=comma, x=epochs, y=Decay] {Fig8.csv};
		\addplot [very thick, black] table[col sep=comma, x=epochs, y expr=1e-5] {Fig8.csv};
		\end{axis}
	
	\begin{axis}[
		xlabel={Epochs},
		ylabel=\textcolor{RougeLMS}{Modes},
		axis y line*=right,
		axis x line=none,
		ytick style={color=RougeLMS},
		yticklabel style={color=RougeLMS},
		ytick={1,2,3},
		]
		\addplot[
		color=RougeLMS,
		very thick,
		]
		table [col sep=comma, x=epochs, y=Modes] {Fig8.csv};

	\end{axis}

\end{tikzpicture}

%% file: Loss_2D_Bi_SVK.tex
\begin{tikzpicture}
	
	\begin{axis}[
		xlabel={Epochs},
		ylabel=\textcolor{BleuLMS}{Loss},
	y dir=reverse,
		ymode=log,
		axis y line*=left,
		xmajorgrids=true,
		ymajorgrids=true,
		ytick style={color=BleuLMS},
		yticklabel style={anchor=east,color=BleuLMS},
		ytick={1e-07,1e-06,1e-05,1e-04,1e-03,1e-02,1e-01,1},
		yticklabels={
			$-\mathdefault{10^{-7}}$,
$-\mathdefault{10^{-6}}$,
$-\mathdefault{10^{-5}}$,
$-\mathdefault{10^{-4}}$,
$-\mathdefault{10^{-3}}$,
$-\mathdefault{10^{-2}}$,
$-\mathdefault{10^{-1}}$,
$-\mathdefault{10^{0}}$
		},
		restrict x to domain=0:3000,
		filter discard warning=false,
		]
		\addplot[
		color=BleuLMS,
		very thick,
		]
		table [col sep=comma, x=epochs, y=Loss] {FigSVK.csv};
		
		\end{axis}
	
	\begin{axis}[
		xlabel={Epochs},
		ylabel=\textcolor{RougeLMS}{Modes},
		axis y line*=right,
		axis x line=none,
		ytick style={color=RougeLMS},
		yticklabel style={color=RougeLMS},
		ytick={1,2,3,4,5,6,7},
		restrict x to domain=0:3000,
		filter discard warning=false,
		]
		\addplot[
		color=RougeLMS,
		very thick,
		]
		table [col sep=comma, x=epochs, y=Modes] {FigSVK.csv};
		
	\end{axis}
	
\end{tikzpicture}

%% file: Loss_2D_Bi_zoom_SVK.tex
\begin{tikzpicture}
	
	\begin{axis}[
		xlabel={Epochs},
		ylabel=\textcolor{BleuLMS}{Loss},
		y dir=reverse,
		ymode=log,
		axis y line*=left,
		xmajorgrids=true,
		ymajorgrids=true,
		ytick style={color=BleuLMS},
		yticklabel style={anchor=east,color=BleuLMS},
		ytick={0.00009,0.00012},
		yticklabels={
  $\mathdefault{-9\mathrm{e}{-5}}$,
$\mathdefault{-1.2\mathrm{e}{-4}}$
		},
		ymin=0.00009, ymax=0.000128,
		filter discard warning=false,
		]
		\addplot[
		color=BleuLMS,
		very thick,
		]
		table [col sep=comma, x=epochs_truncated, y=Loss_truncated] {FigSVK.csv};
		
	\end{axis}
	
	\begin{axis}[
		xlabel={Epochs},
		ylabel=\textcolor{RougeLMS}{Modes},
		axis y line*=right,
		axis x line=none,
		ytick style={color=RougeLMS},
		yticklabel style={color=RougeLMS},
		ytick={1,2,3,4,5,6,7},
		ymajorgrids=true,
		restrict y to domain=0:100,
		filter discard warning=false,
		]
		\addplot[
		color=RougeLMS,
		very thick,
		]
		table [col sep=comma, x=epochs_truncated, y=Modes_truncated] {FigSVK.csv};
		
	\end{axis}
	
\end{tikzpicture}

%% file: LossDecay_2D_Bi_SVK.tex
\begin{tikzpicture}
	
	\begin{axis}[
		xlabel={Epochs},
		ylabel=\textcolor{BleuLMS}{Loss decay},
		ymode=log,
		axis y line*=left,
		xmajorgrids=true,
		ymajorgrids=true,
		ytick style={color=BleuLMS},
		yticklabel style={anchor=east,color=BleuLMS},
		ytick={1e-07,1e-06,1e-05,1e-04,1e-03,1e-02,1e-01,1},
yticklabels={
	$\mathdefault{10^{-7}}$,
	$\mathdefault{10^{-6}}$,
	$\mathdefault{10^{-5}}$,
	$\mathdefault{10^{-4}}$,
	$\mathdefault{10^{-3}}$,
	$\mathdefault{10^{-2}}$,
	$\mathdefault{10^{-1}}$,
	$\mathdefault{10^{0}}$
},
		]
		\addplot[
		color=BleuLMS,
		very thick,
		]
		table [col sep=comma, x=epochs, y=Decay] {FigSVK.csv};
		\addplot [very thick, black] table[col sep=comma, x=epochs, y expr=1e-5] {FigSVK.csv};
		\end{axis}
	
	\begin{axis}[
		xlabel={Epochs},
		ylabel=\textcolor{RougeLMS}{Modes},
		axis y line*=right,
		axis x line=none,
		ytick style={color=RougeLMS},
		yticklabel style={color=RougeLMS},
		ytick={1,2,3,4,5,6,7},
		]
		\addplot[
		color=RougeLMS,
		very thick,
		]
		table [col sep=comma, x=epochs, y=Modes] {FigSVK.csv};

	\end{axis}

\end{tikzpicture}

%% file: Loss_2D_Bi_SV.tex
\begin{tikzpicture}
	
	\begin{axis}[
		xlabel={Epochs},
		ylabel=\textcolor{BleuLMS}{Loss},
	y dir=reverse,
		ymode=log,
		axis y line*=left,
		xmajorgrids=true,
		ymajorgrids=true,
		ytick style={color=BleuLMS},
		yticklabel style={anchor=east,color=BleuLMS},
		ytick={1e-07,1e-06,1e-05,1e-04,1e-03,1e-02,1e-01,1},
		yticklabels={
			$-\mathdefault{10^{-7}}$,
$-\mathdefault{10^{-6}}$,
$-\mathdefault{10^{-5}}$,
$-\mathdefault{10^{-4}}$,
$-\mathdefault{10^{-3}}$,
$-\mathdefault{10^{-2}}$,
$-\mathdefault{10^{-1}}$,
$-\mathdefault{10^{0}}$
		},
		restrict x to domain=0:3000,
		filter discard warning=false,
		]
		\addplot[
		color=BleuLMS,
		very thick,
		]
		table [col sep=comma, x=epochs, y=Loss] {Fig_SV.csv};
		
		\end{axis}
	
	\begin{axis}[
		xlabel={Epochs},
		ylabel=\textcolor{RougeLMS}{Modes},
		axis y line*=right,
		axis x line=none,
		ytick style={color=RougeLMS},
		yticklabel style={color=RougeLMS},
		ytick={1,2,3,4,5,6,7},
		restrict x to domain=0:3000,
		filter discard warning=false,
		]
		\addplot[
		color=RougeLMS,
		very thick,
		]
		table [col sep=comma, x=epochs, y=Modes] {Fig_SV.csv};
		
	\end{axis}
	
\end{tikzpicture}

%% file: Loss_2D_Bi_zoom_SV.tex
\begin{tikzpicture}
	
	\begin{axis}[
		xlabel={Epochs},
		ylabel=\textcolor{BleuLMS}{Loss},
		y dir=reverse,
		ymode=log,
		axis y line*=left,
		xmajorgrids=true,
		ymajorgrids=true,
		ytick style={color=BleuLMS},
		yticklabel style={anchor=east,color=BleuLMS},
		ytick={0.0019,0.0022},
		yticklabels={
  $\mathdefault{-1.9\mathrm{e}{-3}}$,
$\mathdefault{-2.2\mathrm{e}{-3}}$
		},
		ymin=0.0018, ymax=0.0023,
		filter discard warning=false,
		]
		\addplot[
		color=BleuLMS,
		very thick,
		]
		table [col sep=comma, x=epochs_truncated, y=Loss_truncated] {Fig_SV.csv};
		
	\end{axis}
	
	\begin{axis}[
		xlabel={Epochs},
		ylabel=\textcolor{RougeLMS}{Modes},
		axis y line*=right,
		axis x line=none,
		ytick style={color=RougeLMS},
		yticklabel style={color=RougeLMS},
		ytick={1,2,3,4,5,6,7},
		ymajorgrids=true,
		restrict y to domain=0:100,
		filter discard warning=false,
		]
		\addplot[
		color=RougeLMS,
		very thick,
		]
		table [col sep=comma, x=epochs_truncated, y=Modes_truncated] {Fig_SV.csv};
		
	\end{axis}
	
\end{tikzpicture}

%% file: LossDecay_2D_Bi_SV.tex
\begin{tikzpicture}
	
	\begin{axis}[
		xlabel={Epochs},
		ylabel=\textcolor{BleuLMS}{Loss decay},
		ymode=log,
		axis y line*=left,
		xmajorgrids=true,
		ymajorgrids=true,
		ytick style={color=BleuLMS},
		yticklabel style={anchor=east,color=BleuLMS},
		ytick={1e-07,1e-06,1e-05,1e-04,1e-03,1e-02,1e-01,1},
yticklabels={
	$\mathdefault{10^{-7}}$,
	$\mathdefault{10^{-6}}$,
	$\mathdefault{10^{-5}}$,
	$\mathdefault{10^{-4}}$,
	$\mathdefault{10^{-3}}$,
	$\mathdefault{10^{-2}}$,
	$\mathdefault{10^{-1}}$,
	$\mathdefault{10^{0}}$
},
		]
		\addplot[
		color=BleuLMS,
		very thick,
		]
		table [col sep=comma, x=epochs, y=Decay] {Fig_SV.csv};
		\addplot [very thick, black] table[col sep=comma, x=epochs, y expr=1e-5] {Fig_SV.csv};
		\end{axis}
	
	\begin{axis}[
		xlabel={Epochs},
		ylabel=\textcolor{RougeLMS}{Modes},
		axis y line*=right,
		axis x line=none,
		ytick style={color=RougeLMS},
		yticklabel style={color=RougeLMS},
		ytick={1,2,3,4,5,6,7},
		]
		\addplot[
		color=RougeLMS,
		very thick,
		]
		table [col sep=comma, x=epochs, y=Modes] {Fig_SV.csv};

	\end{axis}

\end{tikzpicture}